\documentclass[]{aastex631}

\usepackage{amsmath}
\usepackage{hyperref}
\usepackage{gensymb}

\usepackage{scalerel}
\usepackage{tikz}
\usetikzlibrary{svg.path}
\definecolor{orcidlogocol}{HTML}{A6CE39}
\tikzset{
  orcidlogo/.pic={
    \fill[orcidlogocol] svg{M256,128c0,70.7-57.3,128-128,128C57.3,256,0,198.7,0,128C0,57.3,57.3,0,128,0C198.7,0,256,57.3,256,128z};
    \fill[white] svg{M86.3,186.2H70.9V79.1h15.4v48.4V186.2z}
                 svg{M108.9,79.1h41.6c39.6,0,57,28.3,57,53.6c0,27.5-21.5,53.6-56.8,53.6h-41.8V79.1z M124.3,172.4h24.5c34.9,0,42.9-26.5,42.9-39.7c0-21.5-13.7-39.7-43.7-39.7h-23.7V172.4z}
                 svg{M88.7,56.8c0,5.5-4.5,10.1-10.1,10.1c-5.6,0-10.1-4.6-10.1-10.1c0-5.6,4.5-10.1,10.1-10.1C84.2,46.7,88.7,51.3,88.7,56.8z};
  }
}
\newcommand\orcidicon[1]{\href{https://orcid.org/#1}{\mbox{\scalerel*{
\begin{tikzpicture}[yscale=-1,transform shape]
\pic{orcidlogo};
\end{tikzpicture}
}{|}}}}

\submitjournal{AAS Journals}
\shorttitle{A Large Catalog of White Dwarf Characteristics}
\shortauthors{Crumpler et al.}

\begin{document}

\title{A Large Catalog of DA White Dwarf Characteristics using SDSS and Gaia Observations}

\correspondingauthor{Nicole R. Crumpler}
\author{Nicole R. Crumpler \orcidicon{0000-0002-8866-4797}}
\email{ncrumpl2@jh.edu}
\altaffiliation{NSF Graduate Research Fellow}
\affiliation{William H. Miller III Department of Physics and Astronomy, Johns Hopkins University, Baltimore, MD 21210, USA}

\author{Vedant Chandra \orcidicon{0000-0002-0572-8012}}
\affiliation{Center for Astrophysics $|$ Harvard \& Smithsonian, 60 Garden St, Cambridge, MA 02138, USA}

\author{Nadia L. Zakamska \orcidicon{0000-0001-6100-6869}}
\affiliation{William H. Miller III Department of Physics and Astronomy, Johns Hopkins University, Baltimore, MD 21210, USA}

\author{Gautham Adamane Pallathadka \orcidicon{0000-0002-5864-1332}}
\affiliation{William H. Miller III Department of Physics and Astronomy, Johns Hopkins University, Baltimore, MD 21210, USA}

\author{Stefan Arseneau \orcidicon{0000-0002-6270-8624}}
\affiliation{William H. Miller III Department of Physics and Astronomy, Johns Hopkins University, Baltimore, MD 21210, USA}
\affiliation{Department of Astronomy \& Institute for Astrophysical Research, Boston University, 725 Commonwealth Avenue, Boston, MA, 02215, USA}

\author{Nicola Gentile Fusillo \orcidicon{0000-0002-6428-4378}}
\affiliation{Department of Physics, Universita’ degli Studi di Trieste, Via A. Valerio 2, 34127, Trieste, Italy}

\author{J.J. Hermes \orcidicon{0000-0001-5941-2286}}
\affiliation{Department of Astronomy \& Institute for Astrophysical Research, Boston University, 725 Commonwealth Avenue, Boston, MA, 02215, USA}

\author{Carles Badenes \orcidicon{0000-0003-3494-343X}}
\affiliation{Department of Physics and Astronomy, University of Pittsburgh, 3941 O'Hara Street, Pittsburgh, PA 15260, USA}
\affiliation{Pittsburgh Particle Physics, Astrophysics, and Cosmology Center (PITT PACC), University of Pittsburgh, Pittsburgh, PA 15260, USA}

\author{Priyanka Chakraborty \orcidicon{0000-0002-4469-2518}}
\affiliation{Center for Astrophysics $|$ Harvard \& Smithsonian, 60 Garden St, Cambridge, MA 02138, USA}

\author{Boris T. G{\"a}nsicke \orcidicon{0000-0002-2761-3005}}
\affiliation{Department of Physics, University of Warwick, Coventry CV4 7AL, UK}

\author{Sean Morrison\orcidicon{0000-0002-6770-2627}}
\affiliation{Department of Astronomy, University of Illinois at Urbana-Champaign, Urbana, IL 61801, USA}

\author{Hans-Walter Rix\orcidicon{0000-0003-4996-9069}}
\affiliation{Max-Planck-Institut für Astronomie, Konigstuhl 17, D-69117 Heidelberg, Germany}

\author{Stephen P.\ Schmidt \orcidicon{0000-0001-8510-7365}}
\altaffiliation{NSF Graduate Research Fellow}
\affiliation{William H. Miller III Department of Physics and Astronomy, Johns Hopkins University, Baltimore, MD 21210, USA}

\author{Axel Schwope\orcidicon{0000-0003-3441-9355}}
\affiliation{Leibniz-Institut für Astrophysik Potsdam (AIP), An der Sternwarte 16, 14482 Potsdam, Germany}

\author{Keivan G.\ Stassun\orcidicon{0000-0002-3481-9052}}
\affiliation{Department of Physics and Astronomy, Vanderbilt University, Nashville, TN 37235, USA}

\date{\today}

\begin{abstract}
We present a catalog of 8545 and 19,257 unique DA white dwarfs observed in SDSS Data Release 19 and previous SDSS data releases, respectively. This is the largest catalog of both spectroscopic and photometric measurements of DA white dwarfs available to date, and we make this catalog and all code used to create it publicly available. We measure the apparent radial velocity, spectroscopic effective temperature and surface gravity, and photometric effective temperature and radius for all objects in our catalog. We validate our measurements against other published white dwarf catalogs. For apparent radial velocities, surface gravities, and effective temperatures measured from spectra with signal-to-noise ratios $>50$, our measurements agree with  published SDSS white dwarf catalogs to within 7.5 km/s, 0.060 dex, and 2.4\%, respectively. For radii and effective temperatures measured with Gaia photometry, our measurements agree with other published Gaia datasets to within $0.0005$ $R_\odot$ and $3\%$, respectively. We use this catalog to investigate systematic discrepancies between white dwarfs observed in SDSS-V and previous generations of SDSS. For objects observed in both SDSS-V and previous generations, we uncover systematic differences between measured spectroscopic parameters depending on which set of survey data is used. On average, the measured apparent radial velocity of a DA white dwarf is $11.5$ km/s larger and the surface gravity is $0.015$ dex smaller when a white dwarf's spectroscopic parameters are measured using SDSS-V data compared to using data from previous generations of SDSS. These differences may be due to changes in the wavelength solution across survey generations.
\end{abstract}


\section{Introduction} \label{sec:intro}

\indent When a $\sim0.07-8$ $M_\odot$ star runs out of fuel for fusion, it expels its outer layers, forming a planetary nebula with a hot compact object at its center \citep{Fontaine_2001}. This object is the slowly cooling core of the star, known as a white dwarf (WD). WDs have been observed with a variety of masses ($\sim 0.2-1.3$ $M_\odot$), radii ($0.008-0.2$ $R_\odot$), and atmospheric compositions \citep{Saumon_2022}. WDs with hydrogen-dominated atmospheres, so-called DA WDs, comprise $\sim80$\% of all observed WDs in magnitude-limited samples \citep{Kepler_2019}. By astrophysical standards, DA WDs are relatively simple and well-understood objects, so much so that they are often used for flux calibration \citep{Bohlin_2014}. This solid theoretical understanding enables precise measurements of DA WD stellar parameters from combining spectroscopic and photometric observations with model spectra. Generally, DA WD effective temperatures, surface gravities, and radii can be measured to $5-10$\%, $0.1$ dex, and $0.001-0.002$ $R_\odot$, respectively \citep{Chandra_2020_2, Chandra_2020_1}.

\indent Spectroscopic observations of DA WDs can be used to measure the effective temperature ($T_{\text{eff}}$) and the logarithm of the surface gravity ($\log{g}$) of these stars by fitting model spectra to the observed Balmer series hydrogen absorption lines \citep{Bergeron_1992}. Because of the strong dependence on the assumptions underlying the models used in the spectral fits, improvements in model spectra have increased the accuracy of measurements with this method. Some of these improvements include better characterizations of non-ideal gas effects \citep{Tremblay_2009}, more detailed treatments of collisional broadening \citep{Blouin_2017}, and the incorporation of three-dimensional convective effects \citep{Tremblay_2013}. The impact of three-dimensional effects is a particularly illustrative example of the power of improvements to model spectra, as their inclusion resolved the high $\log{g}$ problem of spectroscopic measurements. The high $\log{g}$ problem was an unphysical increase in the spectroscopic surface gravities (or, correspondingly, the masses) of WDs with spectroscopic effective temperatures of $\sim11,000-13,000$ K. At effective temperatures below $\sim14,000$ K (for a surface gravity of 8 dex), pure hydrogen atmospheres become convective due to an opacity increase from the recombination of hydrogen \citep{Tremblay_2013}. One-dimensional models cannot account fully for these effects, and three-dimensional models resolved this issue by including convective overshoot and other effects \citep{Tremblay_2013}, dramatically improving spectroscopic measurements in $\sim11,000-13,000$ K temperature range.

\indent Photometric observations of WDs can be combined with distance measurements to obtain the radius ($R$) and the effective temperature of the star by fitting these observations to model photometry \citep{Koester_1979}. Historically, this method was limited by the dearth of confident parallax measurements for most WDs \citep{Bergeron_2019}, but precise parallax measurements from the European Space Agency's Gaia Mission \citep{Lindegren_2021}  have now enabled widespread use of photometric measurements of WD properties \citep{Gentile_2021}. Measurements of spectroscopic and photometric temperatures can be directly compared, and have been found to agree to within $2$\% across a broad range of temperatures for isolated DA WDs \citep{Tremblay_2019_wdtools}. Surface gravities from spectra and radii from photometry can also be compared using theoretical mass-radius relations such as the \citet{Fontaine_2001} models. \citet{Tremblay_2019_wdtools} found agreement between surface gravities derived from photometric radii and atmospheric models and spectroscopic surface gravities to within 0.2 dex.

\indent In the modern era of astrophysical observation, missions like Gaia \citep{Gaia_2016, Gaia_2023} and the Sloan Digital Sky Survey \citep[SDSS,][Kollmeier, J.A., et al. 2025, AJ, submitted]{York_2000,Eisenstein_2011,Blanton_2017} have enabled precision photometry and spectroscopy of tens of thousands of DA WDs \citep{Kepler_2019, Gentile_2021}. Large catalogs of WD stellar parameters have opened the door for tests of the detailed underlying physical effects impacting these stars. One such discovery was the possible detection of latent heat released by crystallization, a process by which the WD solidifies as it cools \citep{vanHorn_1968}, in the Q branch of the Gaia WD Hertzsprung-Russell diagram \citep{Tremblay_2019_crystallization}. Further investigation into this effect revealed that some high mass WDs experience an extra cooling delay that cannot be explained by crystallization alone \citep{Cheng_2019}, which was later shown to be due to a distillation process whereby $^{22}$Ne is transported to the center of the WD \citep{Blouin_2021}. Additionally, \citet{Chandra_2020_1} used a large catalog of DA WDs to empirically test the WD mass-radius relation across a wide range of masses using gravitational redshifts, finding good agreement with \citet{Fontaine_2001} models of this relation. \citet{Crumpler_2024} extended this analysis and, using the catalog constructed in this work, directly detected the temperature dependence of the mass-radius relation. With even larger catalogs and higher quality data, this technique may be able to test agreement between model mass-radius relations and empirical measurements as a function of temperature, potentially uncovering yet unknown subtleties in the physics underlying this relation.

\indent As theory and observation continue to improve, even higher order effects on WD structure and evolution may be uncovered and characterized, establishing DA WDs as some of the best understood astrophysical objects. This has important implications for other areas of research, especially searches for new fundamental physics such as revealing the nature of dark matter. WDs are old and dense, providing physical conditions not present on Earth \citep{Saumon_2022}. These stars have already proven to be crucial sources of mass and interaction strength constraints on various dark matter models. Many studies have investigated the cooling or heating effects of different types of dark matter on WDs, and how this might impact the WD luminosity function \citep{Raffelt_1986, Isern_2008,Isern_2010, Althaus_2011, Dreiner_2013, Isern_2018} or the pulsational periods of variable WDs \citep{Isern_1992, Corsico_2001, Benvenuto_2004, Biesiada_2004, Bischoff_2008, Corsico_2012_1, Corsico_2012_2}. Another common approach is to investigate how dark matter might trigger a type Ia supernova, and place constraints using the observed frequency of these explosions \citep{Graham_2015, Bramante_2015, Graham_2018, Acevedo_2019}. Other approaches to using WDs in fundamental physics searches include studying the effects of axions on the WD mass-radius relation \citep{Balkin_2024}, investigating how dark matter capture might form small black holes \citep{Steigerwald_2022} or compact cores \citep{Leung_2013} in the center of WDs, placing constraints using very cold WDs in globular clusters \citep{Bertone_2008, McCullough_2010, Bell_2021}, and many more. Advancements in WD science can produce new or updated constraints on different dark matter models, making such developments vital for progress in fundamental physics.

\indent In this paper, we produce a catalog of measured apparent radial velocities, effective temperatures, surface gravities, and radii for all SDSS DA WDs observed in SDSS Data Releases 1 through 19 with clean Gaia photometry. This catalog contains 8545 and 19,257 unique WDs observed in the 19th Data Release of the on-going SDSS-V survey and in previous generations of SDSS, respectively, making it the largest such catalog to date. In Sec. \ref{sec:catalog}, we describe the process of creating our WD catalogs using data from SDSS-V and previous generations of SDSS. In Secs. \ref{sec:SDSSV_sample} and \ref{sec:eSDSS_sample}, we recount the SDSS survey details, the criteria for selecting DA WDs, and the process for obtaining the WD spectra and photometry for SDSS-V and previous generations of SDSS. In Sec. \ref{sec:rv}, we explain our measurement procedures for fitting WD apparent radial velocities from spectra, and use these measurements along with other methods to flag binary contaminants. In Sec. \ref{sec:logg_teff}, we describe our measurement procedures for fitting WD surface gravities and effective temperatures from spectra. In Sec. \ref{sec:r_teff}, we detail our measurement procedure for fitting WD radii and effective temperatures from photometry. We describe our measured DA WD mass distributions, flag which stellar population each star belongs to, and characterize the completeness of our sample in Sec. \ref{sec:other}. In Sec. \ref{sec:SDSS_diffs}, we detail the systematic differences between DA WDs observed in SDSS-V and those observed in previous generations of SDSS. We discuss the implications of our work in Sec. \ref{sec:conc}. All spectra are on the vacuum wavelength scale. Surface gravities are measured on the $\log{g}$ scale in dex where $g$, the surface gravity, is in CGS units. The code used for all measurements and results found in this paper is publicly available\footnote{\url {https://github.com/nicolecrumpler0230/WDparams}}.

\section{Catalog Construction} \label{sec:catalog}

\indent In this section, we detail the selection criteria and measurement processes used in constructing our two catalogs. The first catalog consists of 8545 unique DA WDs identified in SDSS Data Release 19, part of the on-going SDSS-V survey. The second is made up of 19,257 unique DA WDs identified across previous SDSS data releases. In order to emphasize the impact of the latest generation of SDSS, we keep these two catalogs separate. There is an overlap of 1761 WDs found both in our SDSS-V and previous SDSS catalogs. The SDSS-V and previous SDSS catalogs are publicly available\footnote{\url {https://www.sdss.org/dr19/data_access/value-added-catalogs/?vac_id=10008}} as official SDSS Value Added Catalogs (VACs).

\subsection{SDSS-V Catalog Sample Selection} \label{sec:SDSSV_sample}

\indent The SDSS-V survey (Kollmeier, J.A., et al. 2025, AJ, submitted) began operations in November 2020. Initially, the survey operated only from the 2.5 m telescope located at the Apache Point Observatory in Sunspot, New Mexico \citep{Gunn_2006}, but then expanded to the 2.5 m telescope at Las Campanas Observatory in the Atacama Desert of Chile \citep{Bowen_1973} in 2022 to provide coverage of both the Northern and Southern Hemispheres. At the start of SDSS-V, Apache Point Observatory objects were observed with traditional plug-plates, aluminum plates with holes drilled for each fiber. In 2022, the Apache Point Observatory telescope switched to a robotic focal plane system, which can quickly reconfigure the fibers to accommodate different field observations \citep{Pogge_2020}. The Las Campanas Observatory telescope always operated using the robotic focal plane system. WDs observed in SDSS-V were targeted through the Milky Way Mapper program, which focuses on observing millions of stars in the Galaxy with multi-epoch spectroscopy (Johnson, J.A., et al. in prep). All SDSS-V WD spectra used in this paper were obtained with the Baryon Oscillation Spectroscopic Survey spectrograph \citep[BOSS, ][]{Smee_2013}, which is a low-resolution spectrograph ($R\sim$1,800) covering wavelengths in the range 3,650 - 9,500 \AA. The SDSS-V spectra were reduced using v6\_1\_3 of the BOSS Data Reduction Pipeline, which accounts for a variety of instrumental and atmospheric effects and uses skylines as the radial velocity standard in order to obtain the wavelength solution for each spectrum \citep{Dawson_2013}. By investigating the standard deviation of measured stellar radial velocities, \citet{Adelman_2008} found that the BOSS absolute wavelength calibration is accurate to $\sim 7$ km s$^{-1}$, at the design limit of the spectrograph \citep{Stefan_2023}. 

\indent The SDSS-V catalog is selected from all SDSS-V Data Release 19 objects identified as DA WDs by the spectral classification algorithm included in the SDSS-V pipeline known as \texttt{SnowWhite} as of November 2023. This algorithm classifies WD spectra into 26 possible WD subclasses. It uses a random forest classifier trained on $\sim 30,000$ SDSS WD spectra from SDSS data releases up to 14 classified by visual inspection. Before SDSS-V, the serendipitous WD targeting strategy adopted by SDSS resulted in few observations of WDs with effective temperatures $\lesssim 7000$ K. These cool WDs are therefore significantly underrepresented in the \texttt{SnowWhite} training sample. In order to mitigate the impact of this bias, the training sample of \texttt{SnowWhite} was expanded with the inclusion of 3000 model DA WD spectra from \citet{Tremblay_2013} corresponding to effective temperatures in the range $3000<T_\text{eff}<7000$ K and spanning surface gravities of $7<\log{g}<9.5$ dex. The spectral resolution of the models was degraded to match that of SDSS and random noise was added to create SDSS-like spectra with signal-to-noise ratios (SNRs) varying between 5 and 30. Classification retrieval tests performed on a sample of $\sim$3000 spectra with known classifications excluded from the training set indicated that \texttt{SnowWhite} is 95\% reliable in identifying DA WDs with SNRs$\geq10$. For this work we only select WDs with a reliable DA classification and do not include objects classified as uncertain DAs (denoted by ``DA:" or by dual classification e.g. ``DA/DBA"). We also exclude DA WDs which were classified as members of WD-main sequence star binary systems by \texttt{SnowWhite} (denoted by ``DA\_MS").

\begin{figure}[h!]
\begin{center}
\includegraphics[scale=0.3]{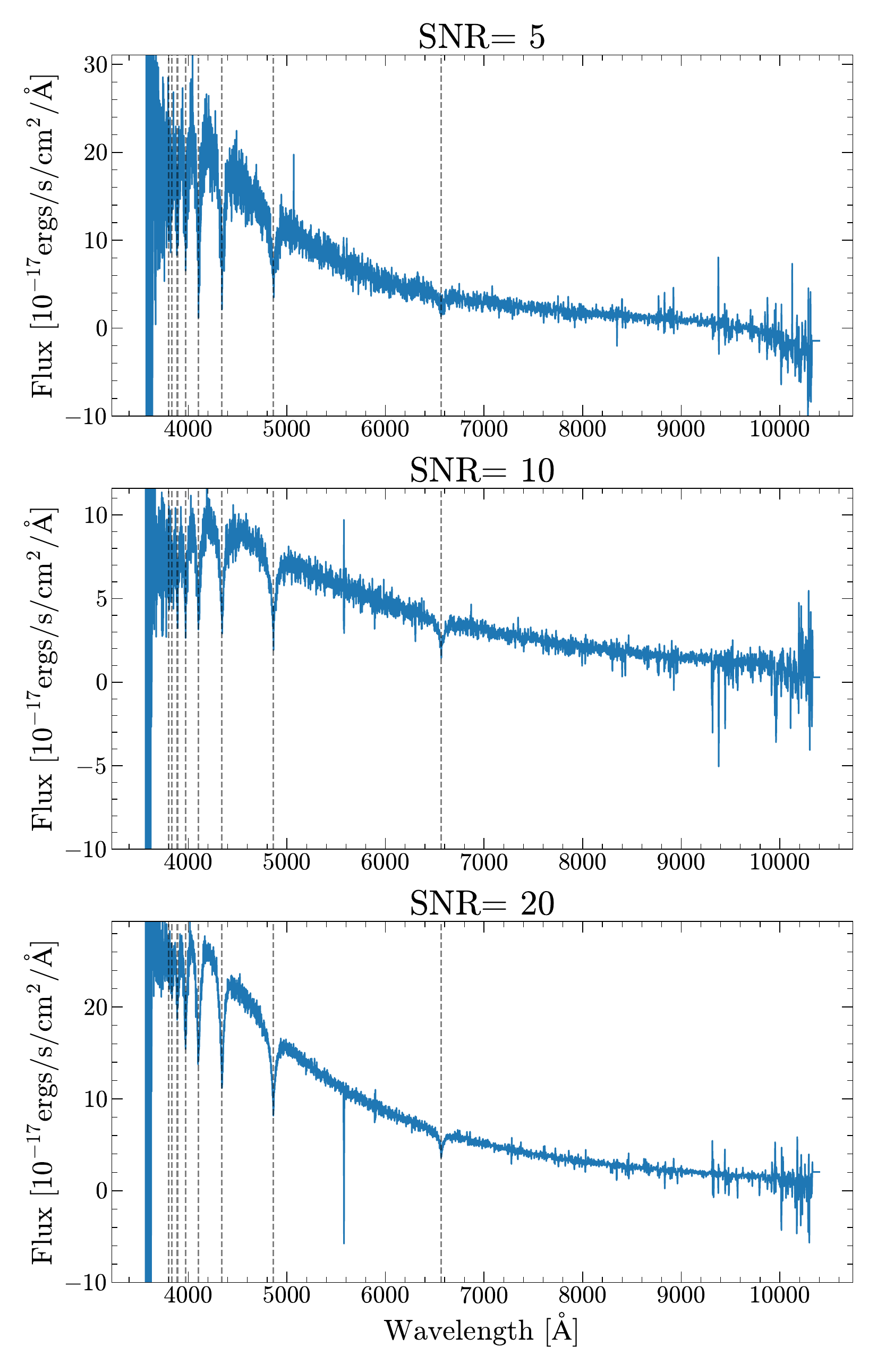}
\caption{Three DA WD spectra from the SDSS-V survey included in this catalog. The top spectrum has a median SNR of 5, the middle has a SNR of 10, and the bottom has a SNR of 20. The dashed lines mark the wavelengths of the hydrogen Balmer series absorption lines, showing that these WDs have hydrogen-dominated atmospheres. \label{fig:DA_spec}}
\end{center}
\end{figure}

\indent Each SDSS-V WD spectrum is uniquely identified by the field ID, Modified Julian Date (MJD), and catalog ID of the observation. Example SDSS-V DA WD spectra with median SNRs of 5, 10, and 20 are shown in Fig. \ref{fig:DA_spec}. Each spectrum in the SDSS-V catalog is obtained by coadding individual 15-minute sub-exposures obtained within the same night or on consecutive nights, with the number of sub-exposures determined by the weather conditions and by how many of them are necessary to build-up the SNR required by the survey.  In this work, we use only the coadded spectra from the SDSS pipeline which we refer to as ``individual observations" or ``individual spectra" throughout the paper. We do not use the 15-minute sub-exposures comprising these coadds. Each WD is uniquely identified by the Gaia DR3 source ID and SDSS-ID included in the SDSS-V pipeline output. We cross-reference these SDSS observations of DA WDs with the Gaia DR3 catalog to obtain the Gaia photometry and proper motion measurements for each WD and the \citet{BailerJones_2021} geometric distances of each object.

\indent We also cross-reference all selected SDSS objects with the SDSS Data Release 17 \citep{Abdurrouf_2022} photometric catalog to obtain the SDSS \textit{ugriz} photometry. We convert SDSS magnitudes to the AB magnitude system using the following conversions
\begin{align}\label{eqn}
    u_{\text{AB}}&=u_{\text{SDSS}}-0.040\\
    g_{\text{AB}}&=g_{\text{SDSS}}\\
    r_{\text{AB}}&=r_{\text{SDSS}}\\
    i_{\text{AB}}&=i_{\text{SDSS}}+0.015\\
    z_{\text{AB}}&=z_{\text{SDSS}}+0.030,
\end{align}
from \citet{Eisenstein_2006}. For all SDSS AB magnitudes, we include a $0.01$ mag error added in quadrature to the recorded SDSS magnitude error to account for the uncertainty in transforming to the AB system\footnote{\url{https://www.sdss4.org/dr12/algorithms/fluxcal/}}.

\indent We retain only objects with \citet{BailerJones_2021} distances and either clean SDSS or clean Gaia photometry. Our criteria for clean photometry are detailed in Sec. \ref{sec:r_teff}. We identify 8545 individual DA WDs across 16,651 observations from SDSS-V Data Release 19 for use in constructing this catalog. WDs in the catalog have between 1 and 21 observations, with the mean and median WD having 1.9 and 1.0 observations, respectively.

\indent To validate our measurement procedures, we cross-match the DA WDs from Data Release 19 with other published data sets. These data sets include other high-resolution spectroscopic DA WD samples such as \citet{Falcon_2010}\footnote{\url {https://cdsarc.cds.unistra.fr/viz-bin/cat/J/ApJ/712/585}} and \citet{Koester_2009}\footnote{\url {https://cdsarc.cds.unistra.fr/viz-bin/cat/J/A+A/505/441\#/browse}}, other low-resolution spectroscopic DA WD samples such as \citet{Kepler_2019}\footnote{\url {https://cdsarc.cds.unistra.fr/viz-bin/cat/J/MNRAS/486/2169}} and \citet{Anguiano_2017}\footnote{\url {https://cdsarc.cds.unistra.fr/viz-bin/cat/J/MNRAS/469/2102\#/browse}}, and other photometric DA WD samples such as \citet{Raddi_2022}\footnote{\url {https://cdsarc.cds.unistra.fr/viz-bin/cat/J/A+A/658/A22\#/browse}} and \citet{Gentile_2021}\footnote{\url {https://cdsarc.cds.unistra.fr/viz-bin/cat/J/MNRAS/508/3877\#/browse}}.

There are 15 WDs contained both in our SDSS-V data set and the \citet{Falcon_2010} data set, 277 in both our and the \citet{Raddi_2022} data set, 21 in both our and the \citet{Koester_2009} data set, 1523 in both our and the \citet{Kepler_2019} data set, 1567  in both our and the \citet{Anguiano_2017} data set, and 8423 in both our and the \citet{Gentile_2021} data set. For the \citet{Anguiano_2017} and \citet{Kepler_2019} sets, we drop duplicate observations of the same WD, retaining only the observation with the highest SNR. Additionally, we flag two objects in \citet{Anguiano_2017} with apparent radial velocities $>29,900$ km/s as clear outliers, and do not include these objects in our validations. We include the published values from these validation catalogs in our public catalog. Appendix \ref{sec:app1} presents the data model for our SDSS-V Data Release 19 DA WD catalog.

\subsection{Previous SDSS Data Release Catalog Sample Selection} \label{sec:eSDSS_sample}

\indent \citet{Gentile_2021} carried out a careful cross-match of all Gaia EDR3 sources in their catalog with SDSS spectral sources up to Data Release 16 with a 3 arcsecond matching radius and propagated coordinates with Gaia proper motions to account for different epochs. All matching SDSS spectra were visually inspected and a classification is provided in their catalog.
Since the \citet{Gentile_2021} catalog of white dwarfs is based on Gaia EDR3, only those SDSS spectral sources with reliable Gaia matches that passed the quality selection criteria imposed by the authors are included in their catalog. 

\indent For this work we select all spectra classified as DA in \citet{Gentile_2021} and cross-match these observations with previously published SDSS WD catalogs to obtain reference physical properties measured from SDSS spectra, since the \citet{Gentile_2021} measurements used only Gaia photometry. These catalogs include the \citet{Kepler_2019} Data Release 14\footnote{\url {https://cdsarc.cds.unistra.fr/viz-bin/cat/J/MNRAS/486/2169}}, \citet{Kepler_2016} Data Release 12\footnote{\url {https://cdsarc.cds.unistra.fr/viz-bin/cat/J/MNRAS/455/3413}}, \citet{Kepler_2015} Data Release 10\footnote{\url {https://cdsarc.cds.unistra.fr/viz-bin/cat/J/MNRAS/446/4078}}, \citet{Kleinman_2013} Data Release 7\footnote{\url {https://cdsarc.cds.unistra.fr/viz-bin/cat/J/ApJS/204/5}}, \citet{Eisenstein_2006} Data Release 4\footnote{\url {https://cdsarc.cds.unistra.fr/viz-bin/cat/J/ApJS/167/40}}, and \citet{Kleinman_2004} Data Release 1\footnote{\url {https://cdsarc.cds.unistra.fr/viz-bin/cat/J/ApJ/607/426}} papers. These data releases only used the 2.5 m telescope at Apache Point Observatory \citep{Gunn_2006}, and so only cover the Northern Hemisphere. Data Releases 1 through 8 used the original SDSS spectrograph. This low-resolution spectrograph ($R\sim$ 1800) covered a wavelength range of 3800 - 9000 \AA \citep{York_2000}. In Data Release 9, SDSS upgraded to the BOSS spectrograph which has more fibers per plate, a smaller fiber diameter, and a wider wavelength range than the previous spectrograph \citep{York_2000, Smee_2013}. 

\indent Each spectrum is uniquely identified by the plate ID, MJD, and fiber ID. Nearly all of these spectra can be accessed on the SDSS Data Release 18 Science Archive Server\footnote{\url {https://dr18.sdss.org/sas/dr18/spectro/}}. Within this site, there are a variety of sub-directories from which the spectra can be obtained. We include a flag in our online catalog to indicate the folder containing each spectrum. There is one plate, plate ID=10658, included in our catalog for which the directory on the SDSS server is empty\footnote{\url {https://dr18.sdss.org/sas/dr18/spectro/sdss/redux/v5_13_2/spectra/lite/10658/}}. This plate was labeled as `good' and included in SDSS Data Release 16, but was removed from future data releases due to changes in the reduction pipeline causing the plate to be re-labeled as `bad'\footnote{\url {https://www.sdss4.org/dr17/spectro/caveats/\#missingplatedr17}}. However, these observations can still be accessed by passing the plate ID, MJD, and fiber ID to \texttt{SDSS.astroquery}\footnote{\url {https://astroquery.readthedocs.io/en/latest/sdss/sdss.html}}, and a flag on our online catalog indicates this special case.

\indent We use the \citet{Gentile_2021} Gaia Early Data Release 3 coordinates of each object in a 3 arcsecond radius cross-match with Gaia DR3 sources to obtain Gaia DR3 source IDs. As in Sec. \ref{sec:SDSSV_sample}, we also obtain the Gaia photometry, \citet{BailerJones_2021} distances, and SDSS photometry for each of these WDs. We retain only objects with either clean SDSS or clean Gaia photometry and \citet{BailerJones_2021} distances. In total, the previous SDSS catalog contains 19,257 individual DA WDs across 25,415 observations. WDs in the catalog have between 1 and 21 observations, with the mean and median WD having  1.3 and 1.0 observations, respectively. To validate our measurements, we compare our measured values to spectroscopic measurements from the most recent published data release containing the SDSS spectrum and photometric measurements from \citet{Gentile_2021}. All of this information is included in our online previous SDSS DA WD catalog, and Appendix \ref{sec:app2} presents the data model for this catalog.

\subsection{Spectroscopic Apparent Radial Velocity Measurement and Validation} \label{sec:rv}

\indent The apparent radial velocity of a DA WD is given by the observed shift in the centroids of the Balmer absorption lines from the measured centroids of these lines in the vacuum. If $\lambda_{obs}$ is the observed wavelength and $\lambda_0$ is the vacuum wavelength of the centroid, the apparent radial velocity can be calculated as
\begin{equation}\label{eqn}
    v_{\text{app}}=\left(\frac{\lambda_{\text{obs}}-\lambda_0}{\lambda_0}\right)c.
\end{equation}

\begin{figure}[h!]
\begin{center}
\includegraphics[scale=0.32]{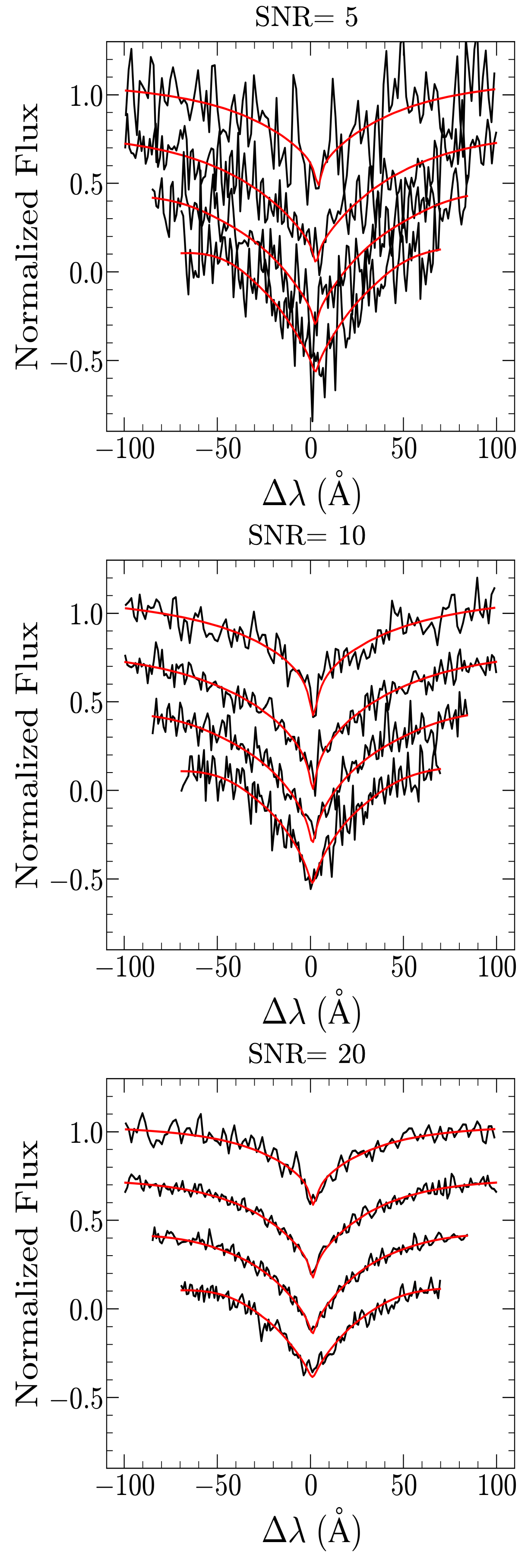}
\caption{\citet{Tremblay_2013} model fits (red) to the spectra (black) shown in Fig. \ref{fig:DA_spec}. From top to bottom, the spectra have SNRs of 5, 10, and 20 and measured apparent radial velocities of $182.6\pm 37.1$, $73.1\pm 18.9$, and $76.2\pm 11.7$ km/s. Within each subplot, from top to bottom, we show the fits to the $H\alpha$, $H\beta$, $H\gamma$, and $H\delta$ hydrogen Balmer series lines. \label{fig:rv_fit}}
\end{center}
\end{figure}

\indent DA WD Balmer lines are apparently broadened by spectral resolution and physically broadened by the Stark effect, which is due to the high pressure in the line-forming areas of the star \citep{Saumon_2022}. These effects necessitate careful treatment when fitting the centroids of Balmer lines, especially considering that the Stark effect creates asymmetric line profiles which can bias apparent radial velocity fits \citep{Napiwotzki_2020}. Model spectra contain a theoretical description of the full shape of the absorption line as opposed to analytic fits, such as Voigt or Gaussian models, which typically fit only the center or the wings of the line. Thus, using model spectra to obtain the apparent radial velocities from spectral lines is more robust than using an analytic fit for spectra with lower SNRs and asymmetric line profiles. For this analysis, we use a publicly available\footnote{\url {https://warwick.ac.uk/fac/sci/physics/research/astro/people/tremblay/modelgrids/}} grid of three-dimensional pure-hydrogen non-local thermodynamic equilibrium (LTE) synthetic spectra from \citet{Tremblay_2013}. Outside the range where three-dimensional convective effects are important, \citet{Tremblay_2013} supplemented this grid with their one-dimensional LTE spectra \citep{Tremblay_2009} at low effective temperatures ($<40,000$ K) and with their one-dimensional non-LTE spectra \citep{Tremblay_2011} at high effective temperatures ($>40,000$ K). Together, these models cover a grid of surface gravities ranging from $7<\log{g}<9$ dex and effective temperatures ranging from $1500<T_{\text{eff}}<130,000$ K. The grid spacing in temperature varies from $250$ K intervals at low temperatures ($1500-5500$ K) to $500$ K intervals at moderate temperatures ($5500-17,000$ K) to even larger intervals ($>3000$ K) at high temperatures ($>17,000$ K). The grid spacing in $\log$ surface gravity is in intervals of $0.5$ dex. Each point in the $\log{g}-T_{\text{eff}}$ grid contains the spectral flux density at the surface of a DA WD in units of erg cm$^{-2}$ s$^{-1}$ Hz$^{-1}$ for air wavelengths ranging between 10 and 94,000 \r{A}\footnote{\url {https://warwick.ac.uk/fac/sci/physics/research/astro/people/tremblay/modelgrids/readme_3d.txt}}. The spectral flux density from the WD's surface is $4\pi$ times the tabulated values from the \citet{Tremblay_2013} models. SDSS wavelengths are given in units of \r{A} for vacuum wavelengths, and the \citet{Tremblay_2013} models are adjusted to match these SDSS spectrum units.

\indent To obtain high-precision apparent radial velocities, we have incorporated the \citet{Tremblay_2013} models into the open-source code Compact Object Radial Velocities \citep[\texttt{corv}\footnote{\url {https://github.com/vedantchandra/corv}},][]{Stefan_2023}. \texttt{corv} fits analytic models or model spectra to observed spectra using $\chi^2$-minimization in order to measure the centroids of Balmer absorption lines and thus the apparent radial velocity of the star. In the fit, all lines are constrained to have the same apparent radial velocity. In fitting the WD apparent radial velocity with \citet{Tremblay_2013} models, \texttt{corv} also returns a measurement of the WD effective temperature and surface gravity. We find that the measurement method employed in Sec. \ref{sec:logg_teff} outperforms the \texttt{corv} routine in measuring the temperature and surface gravity when comparing our measurements to other published data sets. So, although we record the \texttt{corv} temperature and surface gravity measurements in our catalogs, we advise that catalog users utilize the measurements from Sec. \ref{sec:logg_teff} instead. The apparent radial velocities measured from \texttt{corv} are our best-performing apparent radial velocity measurements compared to other data sets.

\indent To understand the biases and uncertainties in our apparent radial velocities, we tested fitting a variety of combinations of the hydrogen Balmer lines for measuring apparent radial velocities. By comparing our measurements to other data sets, we find that, when fitting SDSS spectra with \citet{Tremblay_2013} models, the best results come from fitting the first four hydrogen Balmer lines, that is the $H\alpha$, $H\beta$, $H\gamma$, and $H\delta$ lines. We exclude the higher order lines which are too close together to be fit accurately given the spectral resolution of SDSS. Consequently, all apparent radial velocity fits in our catalog are measured just from the first four hydrogen Balmer lines.

\indent Within both the SDSS-V and previous SDSS sets of observations, we coadd all observations corresponding to each unique WD designated by a unique Gaia DR3 source ID. To coadd the observations, we resample each spectrum so all are on a consistent wavelength grid. We then combine the resulting spectra using a weighted mean, where the weights are determined by the uncertainties on the resampled fluxes. We record the median SNR of each coadded spectrum. We then fit the apparent radial velocities of each individual SDSS spectrum and each coadded spectrum, and record the results. Example fits to the $H\alpha$, $H\beta$, $H\gamma$, and $H\delta$ hydrogen Balmer series lines for the spectra without coadding shown in Fig. \ref{fig:DA_spec} are displayed in Fig. \ref{fig:rv_fit}. 

\subsubsection{Binary System Identification}

\indent We expect that our catalog is contaminated with unresolved double WD binary systems, which can account for 1-10\% of the WD population \citep{Holberg_2009,Toonen_2017,Maoz_2018,Torres_2022}. As the components of a double WD binary system orbit their common center of mass, their observed apparent radial velocities can change over time. Thus, one method to identify such systems is to search for statistically significant variations in the measured apparent radial velocity of a WD across different observations. Following the method of \citet{Maxted_2000} and \citet{Breedt_2017}, we calculate the error-weighted mean apparent radial velocity for each WD having at least two observations with SNR$>10$. We then calculate the $\chi^2$ value of the variation of the measured apparent radial velocities from this mean value, and compare to a $\chi^2$ distribution with $n-1$ degrees of freedom where $n$ is the number of observations of the WD. This gives the probability of observing that amount of apparent radial velocity variation if the source was truly static. \citet{Breedt_2017} take the negative logarithm of this probability and define a parameter $\eta$ that indicates the degree of statistically significant apparent radial velocity variation. As in \citet{Breedt_2017}, we flag any WD with $\eta>2.5$ as a potential binary system. Of the 26,041 unique DA WDs across both the SDSS-V and previous SDSS catalogs, 214 display evidence for binarity from apparent radial velocity variation.

\indent Another method for identifying potential binaries is to search for objects with high Gaia Renormalised Unit Weight Error (RUWE) values. RUWE is a measure of the excess astrometric noise in the system, and values near one indicate sources where a single-star model provides a good fit to the astrometric data. \citet{Sullivan_2025} found that selecting stars with RUWE$>1.2$ is an effective way to identify binary candidates since higher values of RUWE can indicate astrometric noise caused by the motion of a binary system. We flag any WD with RUWE$>1.2$ as a potential binary system. Of the 26,041 unique DA WDs across both the SDSS-V and previous SDSS catalogs, 237 display evidence for binarity from astrometric variation.

\indent 25,593 objects display no evidence for binarity, and only three DA WDs display evidence for binarity from both the $\eta$ and RUWE flags. This is because these binary detection methods probe different regions of binary period parameter space. The $\eta$ method performs best for close double WD binaries, objects with $\log{P}<1-2$ where $P$ is the period of the system in days \citep{Maxted_2000}. Close binaries show stronger apparent radial velocity variation, and are more readily detected via this method. In contrast, the RUWE method requires a sufficiently wide binary system in order to create astrometric noise in the Gaia photometry \citep{Sullivan_2025}. Thus, this method favors double WD systems with longer periods than the $\eta$ method, explaining why few WDs show signs of binarity from both detection methods. In all subsequent validations of our measurement techniques, we include only objects that do not display evidence for binarity.

\indent Binary systems could also be identified by fitting the apparent radial velocities of the the sub-exposures comprising each individual spectrum, but this method is outside the scope of this paper. A forthcoming paper will provide a detailed characterization of all binaries in SDSS-V Data Release 19 (Adamane Pallathadka et al., in prep.), and the results of that paper will supersede the binary flags implemented in this work.

\subsubsection{Validation and Systematic Uncertainties}

\begin{figure}[h!]
\begin{center}
\includegraphics[scale=0.46]{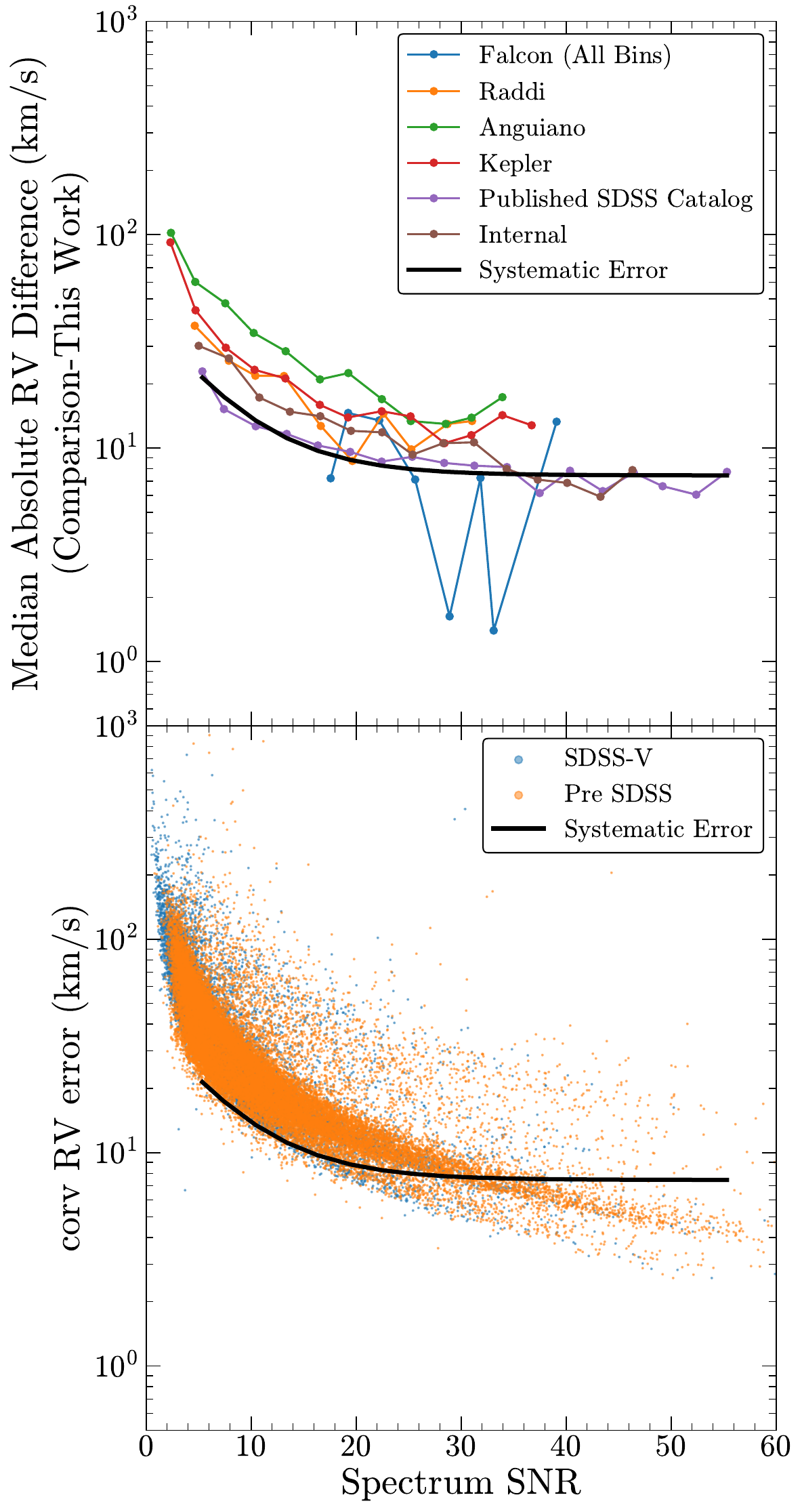}
\caption{The uncertainty in the apparent radial velocity as a function of spectrum SNR. The top plot characterizes the systematic uncertainty in our apparent radial velocity fits by comparing our measurements to other data sets. The blue, orange, green, red, and purple data points show the median absolute difference in each SNR bin when comparing the \texttt{corv} apparent radial velocities to the \citet{Falcon_2010}, \citet{Raddi_2022}, \citet{Anguiano_2017}, \citet{Kepler_2019}, and published SDSS WD catalogs, respectively. The brown data points show an internal comparison. The bottom plot shows the measurement uncertainty returned by \texttt{corv} for the SDSS-V (blue) and the previous SDSS (orange) catalogs. The black curve in both plots is the systematic uncertainty from a decaying exponential fit to the purple points in the top plot, and the functional form of this curve is given in Eqn. \ref{eqn:rv_err}.\label{fig:rv_err_snr}}
\end{center}
\end{figure}

\begin{figure}[h!]
\begin{center}
\includegraphics[scale=0.5]{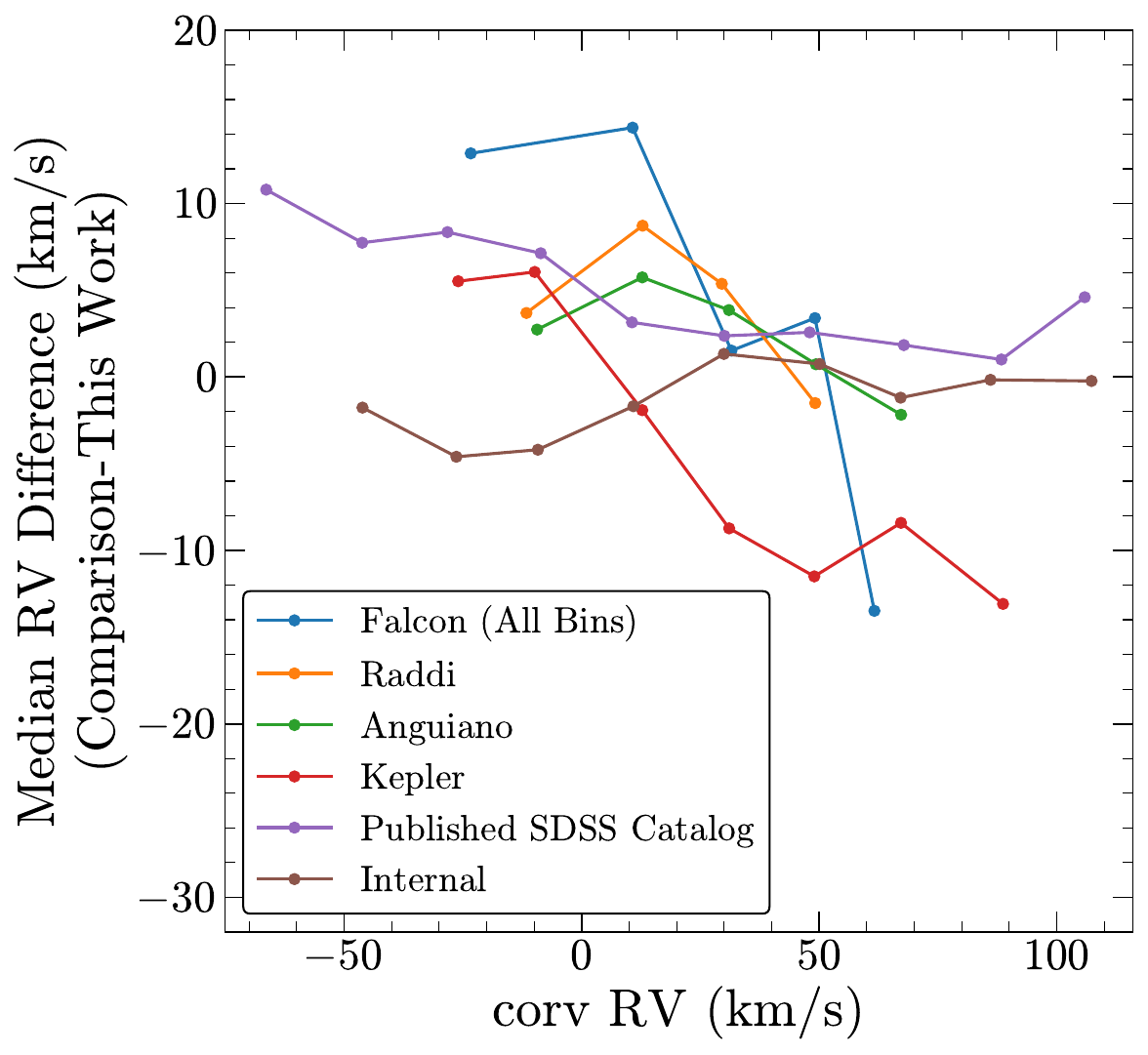}
\caption{The median difference between the apparent radial velocity measured by \texttt{corv} and other datasets as a function of the \texttt{corv} apparent radial velocity, for spectra with SNR$>20$. The blue, orange, green, red, and purple data points show the median difference in each apparent radial velocity bin when comparing the \texttt{corv} apparent radial velocities to the \citet{Falcon_2010}, \citet{Raddi_2022}, \citet{Anguiano_2017}, \citet{Kepler_2019}, and published SDSS WD catalogs, respectively. The brown data points show the same internal comparison as Fig. \ref{fig:rv_err_snr}. \label{fig:rv_err_rv}}
\end{center}
\end{figure}

\indent We validate our fitting method by comparing the measured \texttt{corv} apparent radial velocities to the published velocities in the \citet{Falcon_2010}, \citet{Raddi_2022}, \citet{Anguiano_2017}, and \citet{Kepler_2019} data sets for SDSS-V objects and to published SDSS WD catalog apparent radial velocities for the previous SDSS objects. In all cases, we remove WDs flagged as potential binaries. For both SDSS-V and previous SDSS objects, we compare the apparent radial velocity measured from each unique observation. We plot these comparisons in Figs. \ref{fig:rv_err_snr} and \ref{fig:rv_err_rv}. Fig. \ref{fig:rv_err_snr} shows the median absolute difference in the measured apparent radial velocity as a function of the spectrum SNR for all comparison data sets and for an internal comparison. We choose to use the median absolute difference to characterize agreement between data sets because poor fits due to low SNR or artifacts in the spectrum can cause large outliers which strongly skew other metrics such as the standard deviation. In the internal comparison, we compare our measured apparent radial velocity for a given WD from the highest SNR observation available for that object to other observations with lower SNRs, under the restriction that the highest SNR observation has SNR$\geq20$. The plotted SNR for the internal comparison is the SNR of the lower SNR observation. The overlap with the \citet{Falcon_2010} data set is small, so we include all SNR bins for that catalog. For all other comparisons, we include only bins with at least 20 WDs. For all comparisons and all measurements with a SNR $>20$, the overall median absolute difference between our measured apparent radial velocities and those in published catalogs is within $16$ km/s. For the comparison to published SDSS WD catalogs, the agreement is within $8$ km/s. For the gold-standard \citet{Falcon_2010} data set, which uses high resolution spectra, the agreement is also within $7$ km/s. The absolute wavelength calibration of SDSS-V is $\sim 7$ km/s \citep{Stefan_2023}, meaning that agreement to within $7-8$ km/s is excellent. Our cross-correlation measurement technique approaches the limiting precision of our ability to determine apparent radial velocities with SDSS BOSS spectra. 

\indent The best characterization of the systematic uncertainty in our measurement routine comes from comparing our apparent radial velocities to those from previously published SDSS WD catalogs, since these apparent radial velocities are measured using the same spectra as our previous SDSS catalog. We model our agreement with the SDSS WD catalogs as a decaying exponential and find
\begin{equation}\label{eqn:rv_err}
    \sigma_{\text{RV, sys}}=34e^{-0.2\cdot\text{SNR}}+7.4\text{ km/s},
\end{equation}
where SNR is the median signal-to-noise of the spectrum. This fitted error is plotted in black Fig. \ref{fig:rv_err_snr}, and gives the systematic uncertainty in the measured apparent radial velocity as a function of SNR. A SNR of 5, 10, 20, and 50 results in a systematic uncertainty of 22, 14, 9, and 7.5 km/s respectively. The uncertainty in the measured apparent radial velocity increases dramatically for spectra with SNRs less than 10. Thus, we consider this fitting procedure to be best applied to spectra with a SNR above this threshold. For all individual spectrum and coadded spectrum apparent radial velocity measurements, we add this systematic uncertainty in quadrature to the measurement uncertainty returned by \texttt{corv} to get the full error on the apparent radial velocity measurement. 

\indent In the lower panel of Fig. \ref{fig:rv_err_snr}, we plot the systematic uncertainty of Eqn. \ref{eqn:rv_err} and the measurement error returned by \texttt{corv} as a function of SNR to compare these sources of uncertainty. We emphasize that the plotted systematic uncertainty is not fit to the blue or orange points in the lower plot of Fig. \ref{fig:rv_err_snr}, but is the same curve from the top plot of Fig. \ref{fig:rv_err_snr}, which was fit to the median absolute difference with previously published SDSS WD catalogs (purple points). We find that the systematic uncertainty is typically less than the measurement uncertainty, except at high SNRs. At higher SNR, there are two bands visible in the measurement uncertainty from \texttt{corv} as a function of SNR. The lower band corresponds to nearby objects while the upper band corresponds to more distant objects, with typical distances of $\sim100$ and $\sim200$ pc, respectively. Thus, the band structure results from the fact that, at a fixed SNR, apparent radial velocities are more easily measured for nearby WDs.

\indent We investigate any systematic trends with measured apparent radial velocity in Fig. \ref{fig:rv_err_rv}, which shows the median difference in the measured apparent radial velocity as a function of the measured apparent radial velocity for all comparison data sets and for an internal comparison for spectra with SNR$>20$. For the \citet{Falcon_2010} catalog we include all bins, and for all other comparisons we include only bins with at least 20 entries. We find that all median differences are within $20$ km/s and nearly all are within $10$ km/s. We find no evidence for trends in the discrepancies across the datasets. 

\indent The coadded spectrum apparent radial velocity is one measure of the apparent radial velocity for the object. To validate the coadd measurements, we also calculate a weighted mean apparent radial velocity from the individual spectrum measurements, including only those spectra with a SNR of at least 10. We weigh by the full error on the apparent radial velocity, including the systematic component. The median absolute difference between the apparent radial velocities measured from the coadded spectra and those obtained by taking the weighted mean is $\sim 2$ km/s for coadded spectra with a SNR of at least 10. Thus, either the coadded apparent radial velocities or the weighted mean apparent radial velocities can be used interchangeably for isolated WDs. However, we recommend using the coadded spectrum parameters because some objects may have a coadded spectrum with a SNR $>10$, but have no individual spectrum with SNR $>10$. For these objects, there are well-measured apparent radial velocities from the coadded spectrum but no corresponding mean apparent radial velocity. The median absolute difference between the coadded or weighted mean apparent radial velocities and the values measured from individual observations is $\sim 8-9$ km/s. Again, this is close to the SDSS absolute wavelength calibration, so this agreement is excellent.

\subsection{Spectroscopic Surface Gravity and Effective Temperature Measurement and Validation} \label{sec:logg_teff}

\indent The shapes of the Balmer series lines in DA WD spectra depend on the effective temperature and surface gravity of the star \citep{Tremblay_2013}. To measure these spectroscopic parameters, we employ a modified version of the publicly available code \texttt{wdtools\footnote{\url {https://wdtools.readthedocs.io/en/latest/}}} \citep{Chandra_2020_2}. This code contains a parametric random forest routine, which is trained on the \citet{Tremblay_2019_wdtools} catalog to build a regression between the structure of the absorption lines and the temperature and surface gravity of DA WDs. The \citet{Tremblay_2019_wdtools} spectroscopic parameters are measured using the one-dimensional models of \citet{Tremblay_2011} and adjusted for three-dimensional effects, such as convective overshoot, using the corrections from \citet{Tremblay_2013}. \texttt{wdtools} determines the basic structure of each line using a Voigt profile analytic fit. Voigt profiles cannot fully describe the asymmetric shape of DA WD absorption lines, but the parameters from these fits can be used as inputs into machine learning methods used to predict the labeled temperature and surface gravity from \citet{Tremblay_2019_wdtools}. The parametric random forest regression is bootstrapped using these Voigt fit parameters as inputs to measure the corresponding temperature and surface gravity of the star. We update the \texttt{wdtools} routine to handle fitting the first eight Balmer lines, since the original code can only fit the first four lines, and make this new version available to the public. We test the quality of spectroscopic fits using four, six, seven, and eight lines. We find that this method performs best when using the first six Balmer lines (H$\alpha$, H$\beta$, H$\gamma$, H$\delta$, H$\epsilon$, H$\zeta$), as higher order lines become too noisy resulting in poor fits or causing the fit to fail altogether. This method does not use a grid of model spectra.

\indent In \citet{Chandra_2020_1}, the authors use the \texttt{wdtools} generative fitting pipeline instead. We test this pipeline and find that the generative fitting pipeline produces less accurate results when compared to published data sets, takes far longer to fit spectra, and sometimes fails to fit spectra due to poor continuum normalization. Thus, we determine that the parametric forest routine is better suited for our analysis. As in Sec. \ref{sec:rv}, we fit the spectroscopic surface gravities and temperatures of each individual SDSS spectrum and each coadded spectrum. 

\begin{figure}[h!]
\begin{center}
\includegraphics[scale=0.45]{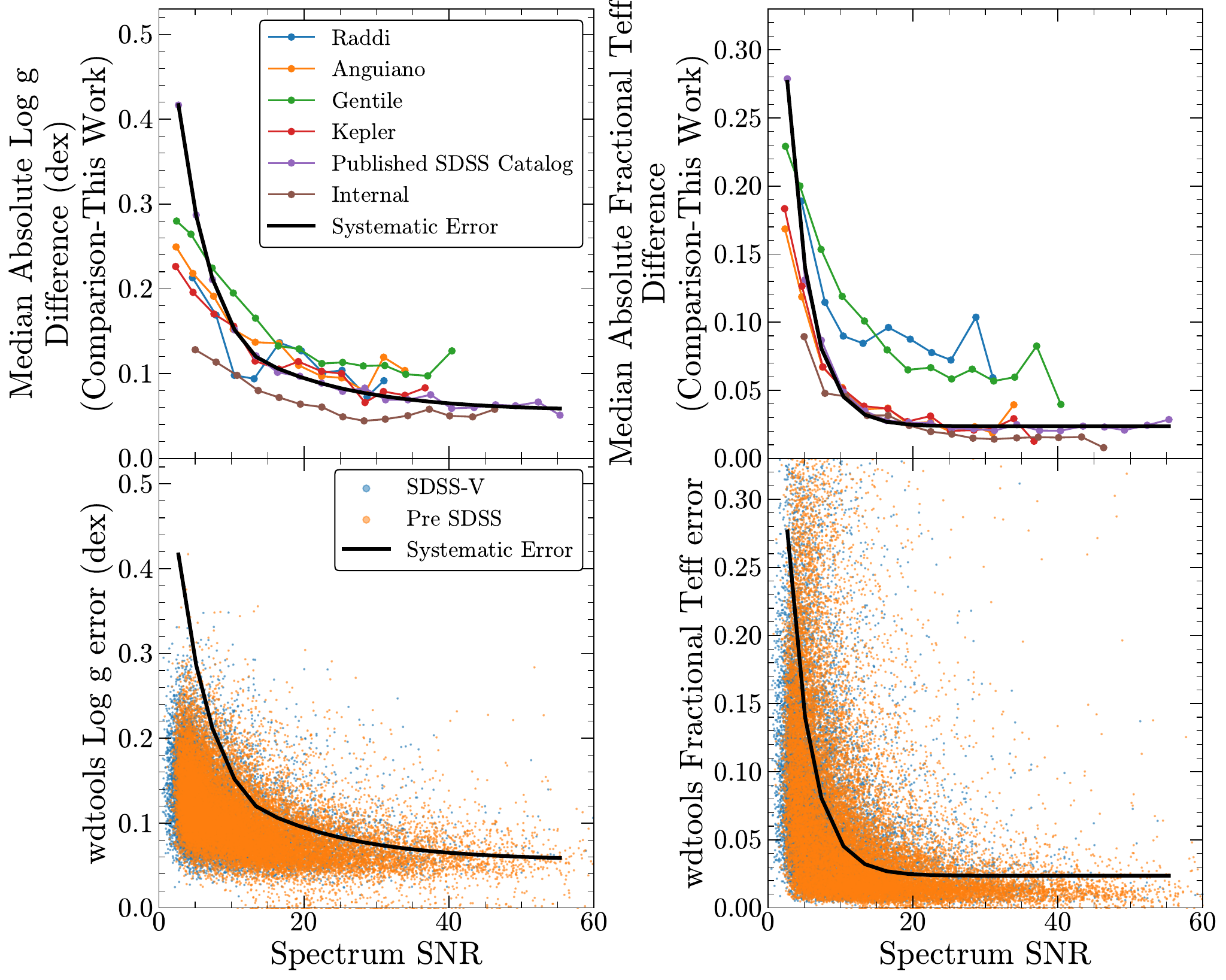}
\caption{The uncertainty on the spectroscopic surface gravity (left) and effective temperature (right) as a function of spectrum SNR. The top plot characterizes the systematic uncertainty in our fitted parameters by comparing our measurements to other data sets. The blue, orange, green, red, and purple data points show the median absolute difference in each SNR bin when comparing the \texttt{wdtools} spectroscopic surface gravity or temperature to the \citet{Raddi_2022}, \citet{Anguiano_2017}, \citet{Gentile_2021}, \citet{Kepler_2019}, and published SDSS WD catalogs, respectively. The brown data points show an internal comparison. The bottom plot shows the measurement uncertainty returned by \texttt{wdtools} for the SDSS-V (blue) and the previous SDSS (orange) catalogs. The black curve in all plots is the systematic uncertainty from a decaying exponential fit to the purple points, and the functional form of this curve is given in Eqns. \ref{eqn:spec_logg_err} and \ref{eqn:spec_teff_err} for the left and right plots, respectively.\label{fig:spec_err_snr}}
\end{center}
\end{figure}

\subsubsection{Validation and Systematic Uncertainties}

\begin{figure}[h!]
\begin{center}
\includegraphics[scale=0.45]{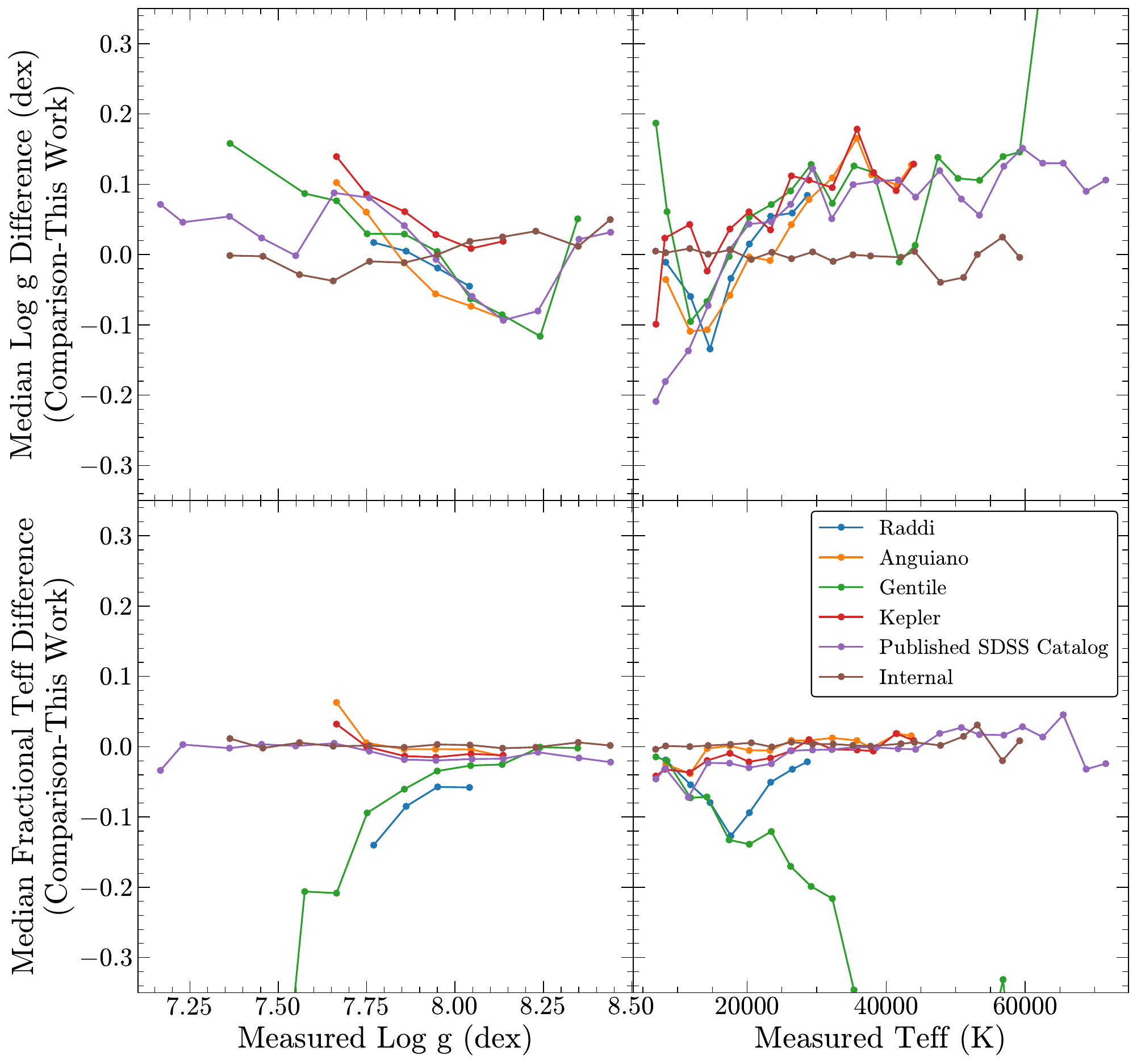}
\caption{The median difference between the surface gravity (top) and median fractional difference between the effective temperature (bottom) measured using \texttt{wdtools} and from other datasets, as a function of our measured surface gravity (left) and temperature (right). The blue, orange, green, red, and purple data points show the median absolute difference in each bin when comparing our measurements to the \citet{Raddi_2022}, \citet{Anguiano_2017}, \citet{Gentile_2021}, \citet{Kepler_2019}, and published SDSS WD catalogs, respectively. The brown data points show the same internal comparison as Fig. \ref{fig:spec_err_snr}. \label{fig:spec_err}}
\end{center}
\end{figure}

\indent We validate our fitting method by comparing the measured \texttt{wdtools} temperatures and surface gravities to the published measurements from the \citet{Raddi_2022}, \citet{Anguiano_2017}, \citet{Gentile_2021}, \citet{Koester_2009}, and \citet{Kepler_2019} data sets for SDSS-V objects and from previous SDSS WD catalogs for the previous SDSS objects. We remove any objects flagged as potential binaries from these comparisons. We plot these comparisons in Figs. \ref{fig:spec_err_snr} and \ref{fig:spec_err}. Fig. \ref{fig:spec_err_snr} displays the median absolute difference in the measured spectroscopic surface gravity and temperature as a function of spectrum SNR for all comparison sets and for the same internal comparison as Sec. \ref{sec:rv}, which compares our measured parameters from the highest SNR$\geq20$ spectrum to those from lower SNR spectra. For all comparisons, we include only bins with at least 20 entries. For all comparison data sets and all observations with a SNR$>20$, the median absolute difference between our measured and the catalog measured effective temperature ranges from $300-1400$ K. In terms of median absolute fractional error, our effective temperature measurements agree with the published catalogs to within $2-8$\%. We consider only SNR$>20$ in these overall comparisons to evaluate the best possible performance of our fitting procedures, as including low SNR spectra would dramatically skew these results. There is particularly good agreement (within $400$ K or $2.5$\%) with the \citet{Anguiano_2017}, \citet{Kepler_2019}, and previously published SDSS data sets. The worst agreement is with the \citet{Raddi_2022} and \citet{Gentile_2021} data sets, with median absolute differences of $1370$ and $980$ K (or $8\%$ and $6\%$), respectively. Both data sets measure the WD surface gravity and temperature from photometry, not spectroscopy, which is the most likely cause of these offsets. Similarly, the median absolute difference between our measured and the catalog measured surface gravity ranges from $0.08-0.13$ dex for observations with SNR$>20$. There is particularly good agreement (within $0.10$ dex) with the \citet{Raddi_2022}, \citet{Anguiano_2017}, \citet{Kepler_2019} and the previous published SDSS data sets. The worst agreement is with the  \citet{Gentile_2021} and \citet{Koester_2009} data sets, with median absolute differences $> 0.11$ dex. For context, effective temperatures are generally calculated to within $500-1000$ K ($\sim 5\%$) and surface gravities to within $0.1$ dex \citep{Chandra_2020_2}. Thus, our fitting techniques are in excellent agreement with the other catalogs and approach the limiting precision of our ability to determine WD spectral parameters with SDSS BOSS spectra and available models. 

\indent As in Sec. \ref{sec:rv}, the best characterization of the systematic uncertainty in our measurements comes from comparing our spectroscopic measurements to those from published SDSS catalogs, which use the same spectra. We model our agreement with the SDSS WD catalogs using decaying exponential fits and find
\begin{align}
    \sigma_{\text{Log g}, sys}&=\left\{\begin{array}{ll}
      0.58e^{-0.20\cdot\text{SNR}}+0.08\text{ dex} & \text{SNR}< 15 \\
      0.16e^{-0.07\cdot\text{SNR}}+0.06\text{ dex} & \text{SNR}\geq 15\\
\end{array} \right.\ \label{eqn:spec_logg_err}\\
    \sigma_{\text{Teff, sys}}&=0.59e^{-0.32\cdot\text{SNR}}+0.02\text{ (fractional)}\label{eqn:spec_teff_err}
\end{align}
where SNR is the median signal-to-noise of the spectrum. These fitted errors are plotted in Fig. \ref{fig:spec_err_snr}, and give the systematic uncertainty in the measured spectroscopic parameters. For the systematic error on the surface gravity, we plot the piecewise fit to the uncertainty as one continuous curve in Fig. \ref{fig:spec_err_snr}. A SNR of 5, 10, 20, and 50 results in a systematic uncertainty of 14.4\%, 4.8\%, 2.5\%, and 2.4\% respectively for the effective temperature and 0.292, 0.158, 0.095, and 0.060 dex respectively for the surface gravity. The uncertainty in the measured spectroscopic parameters increases dramatically for spectra with SNRs less than 10. Thus, as in Sec. \ref{sec:rv}, we consider this fitting procedure to be best applied to spectra with a SNR above this threshold. For all individual spectrum and coadded spectrum surface gravity and effective temperature measurements, we add this systematic uncertainty in quadrature to the measurement uncertainty returned by \texttt{wdtools} to get the full error on these measurements. In Fig. \ref{fig:spec_err_snr}, we also plot the errors returned by \texttt{wdtools} as a function of SNR. We find that these errors are often underestimated relative to the systematic uncertainty, so including the systematic uncertainty is important in fully characterizing the error on the spectroscopic measurements.

\indent In Fig. \ref{fig:spec_err}, we investigate any trends with the measured spectroscopic surface gravity and temperature as a function of the \texttt{wdtools} surface gravity or temperature. For all comparisons, we include only bins with at least 20 entries. We find systematic differences between our effective temperature and those of \citet{Gentile_2021} and \citet{Raddi_2022}, both of which use photometry, not spectroscopy. For comparisons of our spectroscopically-measured temperature to other spectroscopic data sets, we find excellent agreement across the full range of surface gravitites and temperatures with nearly all temperature discrepancies being $<4$\%. For surface gravity measurements, we do not find any evidence for trends in the discrepancies across data sets. For comparison sets measuring surface gravities using spectroscopic methods, nearly all discrepancies are within $0.1$ dex. The data from Fig. \ref{fig:spec_err} is available online\footnote{\url {https://github.com/nicolecrumpler0230/WDparams}}, and can be used to create transformation relations between the surface gravities and effective temperatures measured in this work and those from the comparison catalogs.

\indent The \citet{Tremblay_2019_wdtools} training data set utilized one-dimensional model spectra with three-dimensional corrections to account for the high $\log{g}$ problem in fitting spectroscopic surface gravities \citep{Tremblay_2013}. We investigate our surface gravity measurements as a function of measured temperature and find some structure in the high $\log{g}$ temperature regime. This structure has two components, a $\sim0.2$ dex increase in $\log{g}$ measurements and some clusters within those measurements in the $\sim 11,000$ - $13,000$ K temperature range. The increase in the $\log{g}$ measurements is caused by a residual high $\log{g}$ bump in the \citet{Tremblay_2019_wdtools} training data, which retains this structure despite being corrected for three-dimensional effects. The clusters in the measurements are caused by the three-dimensional corrections and some of the same structure in the \citet{Tremblay_2019_wdtools} training data. The random forest routine is trained only on the spectroscopic information, and does not \textit{a priori} have knowledge of the analytical three-dimensional corrections being applied only in this regime. Because the measured spectral parameters depend only on the shapes of the spectral lines for most of the training data, the random forest routine struggles to encapsulate the effects of the three-dimensional corrections in this temperature window, leading to a clustered structure in the surface gravity fits that is not present outside the $\sim 11,000$ - $13,000$ K temperature range.

\indent For all spectroscopic surface gravity and effective temperature measurements, we also calculate the weighted mean values of these parameters for each object. The weighted means are calculated from all individual spectra associated with a given WD with a SNR of at least 10. The median absolute difference between the effective temperatures and surface gravities measured from the coadded spectra and those obtained by taking the weighted mean is $\sim 1$\% and $0.03$ dex respectively. When compared to spectroscopic parameters measured from individual spectra, the median absolute difference between the coadded or weighted mean effective temperatures and the individually measured values is $\sim 1-2$\%. For surface gravities, the median absolute difference is $\sim 0.04-0.05$ dex. This agreement is excellent relative to the $5\%$ and $0.1$ dex benchmarks \citep{Chandra_2020_2}. Thus, either the coadded or the weighted mean spectral parameters can be used interchangeably. As in Sec. \ref{sec:rv}, we recommend that catalog users utilize the coadded spectroscopic parameters because there are more objects that have a coadded spectrum with SNR$\geq10$. 

\subsection{Photometric Radius and Effective Temperature Measurement and Validation} \label{sec:r_teff}

\indent The \citet{Tremblay_2013} models give the flux at the surface of the WD as a function of wavelength, effective temperature, and surface gravity. These models can be convolved with a photometric filter response curve to give the model flux in that filter at the surface of the WD. Since the flux observed at Earth is the flux at the surface of the star scaled by  $(R/d)^2$ and the distance to the WD is known from the \citet{BailerJones_2021} catalog, we thus obtain the model flux that would be observed at Earth as a function of the WD radius, surface gravity, and effective temperature. We fit these model fluxes to the observed WD photometry. This fitting procedure does not assume a mass-radius relation.

\indent For each object, we fit both clean de-reddened SDSS \textit{urz} photometry and clean de-reddened Gaia $G_{\text{BP}}$ and $G_{\text{RP}}$ photometry if available. SDSS photometry is considered to be clean if the \textit{urz} photometry exists, the \textit{urz} fluxes are $>0$, and the \texttt{clean} photometry flag returned by the SDSS query indicates no issues with the measurements. Fitting all 5 SDSS bands slows the fitting time substantially and does not improve the quality of the fit relative to fitting just the \textit{urz} bands. The median absolute difference in measured radius or effective temperature when fitting with \textit{urz} or \textit{ugriz} bands is orders of magnitude less than the uncertainty on each measurement. Gaia photometry is considered to be clean if the $G_{\text{BP}}$ and $G_{\text{RP}}$ photometry exists and the $G_{\text{BP}}-G_{\text{RP}}$ excess factor is $<2$. Gaia $G$-band magnitudes are nearly degenerate with the $G_{\text{BP}}$ and $G_{\text{RP}}$ magnitudes, and so are not used in the fitting procedure. We recommend that catalog users prioritize SDSS over Gaia photometry, even though Gaia photometry often has smaller errors than SDSS photometry. This is because the \textit{u}-band constraint from SDSS is important for more accurately determining the photometric parameters of the WD, especially at high effective temperatures when the peak of the WD flux becomes bluer. 

\indent To de-redden both the SDSS and Gaia magnitudes, we tested publicly available three-dimensional reddening maps from \citet{Capitanio_2017}, \citet{Green_2019}, \citet{Lallement_2022}, and \citet{Edenhofer_2024}. We found that the \citet{Lallement_2022} and \citet{Edenhofer_2024} produce similar results. The \citet{Capitanio_2017} map is similar to the \citet{Lallement_2022} and \citet{Edenhofer_2024} maps, but has a sharper increase in reddening at small distances. The \citet{Green_2019} map differed from the others. We choose to use the \citet{Edenhofer_2024} dust map from the  \texttt{dustmaps}\footnote{\url {https://dustmaps.readthedocs.io/en/latest/index.html}} Python package \citep{Green_2018} because this map was corroborated by the \citet{Lallement_2022} map and covered a larger range of distances.

\indent The \citet{Edenhofer_2024} reddening map is valid between approximately 69 pc and 1250 pc. For the SDSS-V and previous SDSS catalogs, 95\% of WDs have median geometric \citet{BailerJones_2021} distances between 80 to 2240 pc and 80 to 1720 pc, respectively. For WDs closer than the distance range covered by the \citet{Edenhofer_2024} map, we set the extinction to 0 and assume any reddening is negligible. For the $\sim5\%$ of WDs that are more distant than 1250 pc, we use the reddening at a distance of 1244 pc to obtain a minimum on the reddening for each object. We choose 1244 pc because it is the maximum distance for which the \citet{Edenhofer_2024} map always returns a valid reddening measurement.  Because of this cutoff, our photometric parameters for WDs more distant than 1244 pc should be treated with caution as underestimated reddening can bias the fit towards hotter radius solutions. Following \citet{Edenhofer_2024}, we obtain the Johnson V-band extinction ($A_V$) at the Galactic coordinates ($l,b$) of each WD for the 16th percentile, median, and 84th percentile (i.e. $\pm1\sigma$) geometric distances from \citet{BailerJones_2021}. We use the \texttt{pyphot}\footnote{\url {https://mfouesneau.github.io/pyphot/}} Python package to characterize the SDSS and Gaia filter curves. We then combine the V-band extinctions and filter information with the extinction curve from \citet{Fitzpatrick_1999} using the \texttt{extinction}\footnote{\url {https://extinction.readthedocs.io/en/latest/}} Python package to derive the extinction in magnitudes for each of the SDSS and Gaia photometric bands. We subtract this extinction to obtain the de-reddened magnitudes in all SDSS and Gaia filters for the three WD distances. Finally, we convert the SDSS AB-system \citep{Eisenstein_2006} and the Gaia Vega-system \citep{Riello_2021} magnitudes to physical fluxes.

\begin{figure}[h!]
\begin{center}
\includegraphics[scale=0.45]{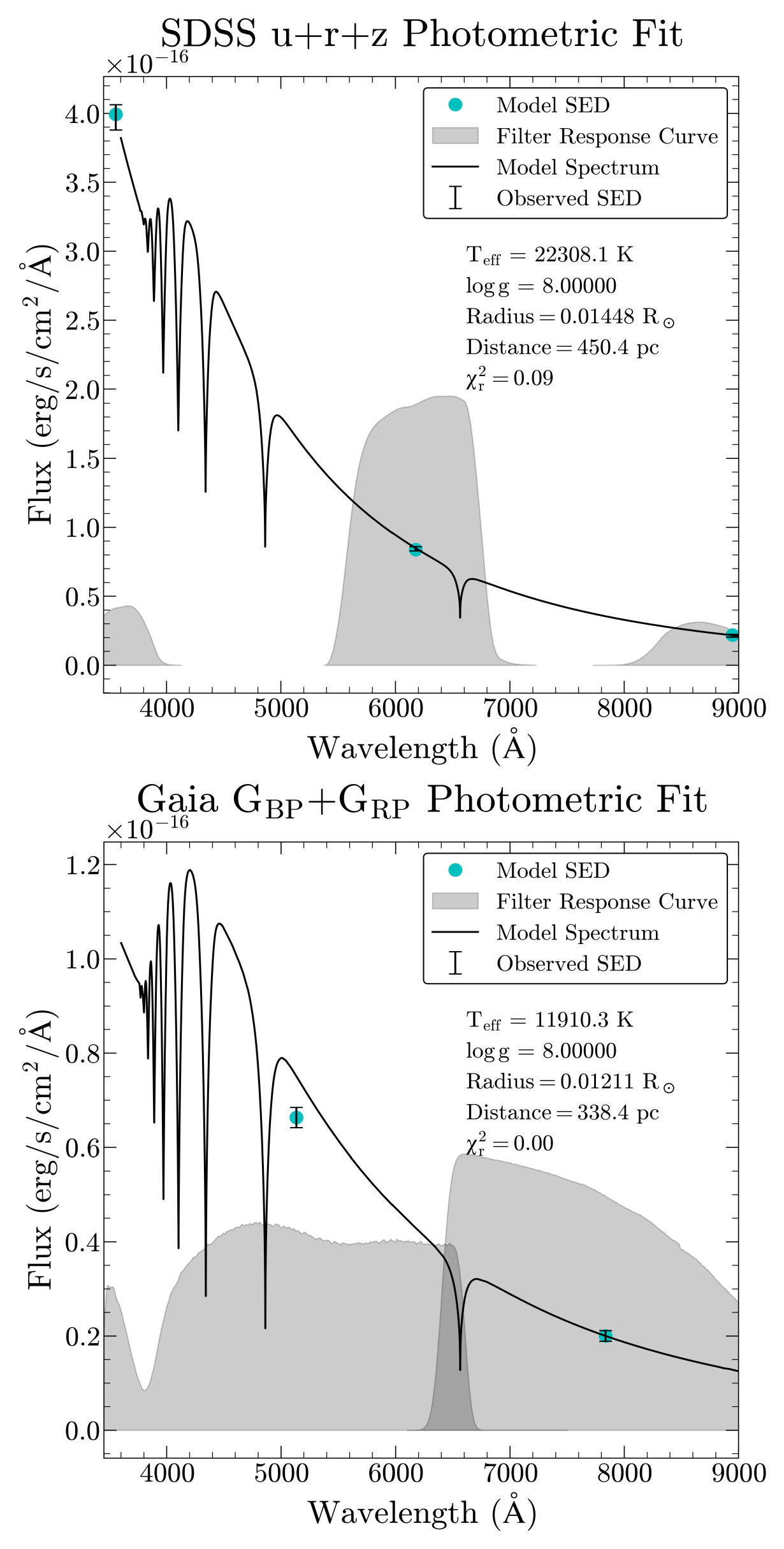}
\caption{Example photometric fits. The top plot displays a fit to the SDSS \textit{urz} bands for the DA WD from Fig. \ref{fig:DA_spec} with a SNR of 20. The bottom plot displays a fit to the Gaia $G_{\text{BP}}$ and $G_{\text{RP}}$ bands for the DA WD from Fig. \ref{fig:DA_spec} with a SNR of 10. The model SDSS or Gaia fluxes are the blue points, from left to right the SDSS \textit{urz} or the Gaia $G_{\text{BP}}$ and $G_{\text{RP}}$ filter response curves are the gray regions, the \citet{Tremblay_2013} model spectrum is the black line, and the observed de-reddened SDSS or Gaia fluxes are the black points with error bars. Some error bars are smaller than the model flux points.\label{fig:phot_fit}}
\end{center}
\end{figure}

\indent With the model and measured photometry in hand, we fit the model fluxes to the observed de-reddened WD photometry via $\chi^2$ minimization to measure the effective temperature and radius of the star. The measured photometry includes the individual uncertainties on the SDSS or Gaia fluxes in each band. Including these uncertainties is particularly important for SDSS photometry, as the error on \textit{u}-band photometry can be much larger than the other bands. For the SDSS fit, we first fit the photometric radius and temperature leaving $\log{g}$ free to vary. If the fit fails or comes within $0.001$ dex of the edge of the surface gravity grid (see Sec. \ref{sec:logg_fixed}), we re-fit the stellar parameters fixing the surface gravity of the WD to 8 dex, the peak value in the WD mass distribution. In this case, we also fit the photometry with surface gravities of $7$ and $9$  dex, and add the difference in measured radius and temperature in quadrature to the radius and temperature errors. We include a flag to indicate which fitting procedure is used for each object. We do not validate these SDSS photometric surface gravities and instead encourage catalog users to utilize the spectroscopic surface gravity measurements of Sec. \ref{sec:logg_teff}, since the shapes of spectral lines better constrain this parameter. For Gaia photometric fits, we fit the radius and temperature fixing the surface gravity to $7$, $8$, and $9$ dex, taking the $8$ dex fit to be the measurement and adding the difference in radius and temperature for the $7$ and $9$ dex fits in quadrature to the errors on the stellar parameters.

\indent An example SDSS photometric fit and Gaia photometric fit for the WDs with a SNR of 20 and 10 from Fig. \ref{fig:DA_spec} are shown in Fig. \ref{fig:phot_fit}. We perform these fits for the photometry corresponding to the 16th percentile, median, and 84th percentile distances. We take the median distance measurements as the measured radius and temperature of the star. We add the difference in temperature or radius for the near and far WD distances in quadrature to the returned measurement errors to account for the uncertainty in the distance to the WD. 

\subsubsection{Effects of Fixed Surface Gravity}\label{sec:logg_fixed}

\indent In \citet{Gentile_2021}, the authors obtain temperatures and surface gravities from Gaia photometry by assuming a mass-radius relation and converting the radius term in the photometric fit to surface gravity. In our method, we retain the dependencies on surface gravity, radius, and temperature to avoid assuming a mass-radius relation. This enables our results to be used to constrain the agreement between photometric measurements and theoretical WD mass-radius relations. 

\indent We test fitting the photometric radius and temperature with leaving the surface gravity free to vary and with fixing the surface gravity to $8$ dex. We find that the surface gravity of WDs is most sensitive to the shapes of the spectral lines, and so does not dramatically alter the observed flux and is not well-constrained by photometry fitted with this method. In the case of Gaia photometry, fitting the photometric surface gravity in addition to the radius and temperature necessitates including Gaia $G$-band photometry. Because Gaia $G$ is nearly degenerate with Gaia $G_{\text{BP}}$ and $G_{\text{RP}}$ measurements, the photometric fit hits the edge of the surface gravity model grid at $7$ or $9$ dex or fails completely in $\sim90\%$ of fits. Thus, Gaia photometry does not effectively constrain photometric surface gravities using this method. In the case of SDSS photometry, we are able to fit photometric surface gravities in $\sim42\%$ of fits. In all other cases the fitting procedure hits the edge of the surface gravity model grid or fails.

\indent We investigate the impact of fixing the surface gravity to 8 dex on other measured parameters by looking at the difference in measured radius and temperature between fits with surface gravity fixed to 7 dex and 9 dex. This comparison represents the worst case scenario, as many of the WDs are likely to be near the sharp peak of the WD mass distribution near $8$ dex. We find that the median difference in radius between the 9 dex and 7 dex fits is $\sim 0.0006$ $R_\odot$, which is small compared to the difference in radius measurements due to the uncertainty in the \cite{BailerJones_2021} distances. In the latter case, the median difference in radius between the 84th percentile and 16th percentile fits is $\sim0.002$ $R_\odot$, an order of magnitude larger than the difference due to fixing the surface gravity. We find that the median difference in temperature between the 7 dex and 9 dex fits is $\sim 1200$ K, which is large compared to the difference in temperature measurements ($\sim 40$ K) due to the uncertainty in the distances. This discrepancy is largely due to objects with effective temperatures of $\sim10,000-15,000$ K, for which the difference in measured temperature can be as large as $20\%$ between the $7$ and $9$ dex fits. Outside this temperature range, the discrepancy is $\sim 5\%$. Thus, we find that fixing the surface gravity during fitting has a negligible effect on the measured radius but may have an important effect on objects with effective temperatures of $\sim 13,000$ K. 

\begin{figure}[h!]
\begin{center}
\includegraphics[scale=0.45]{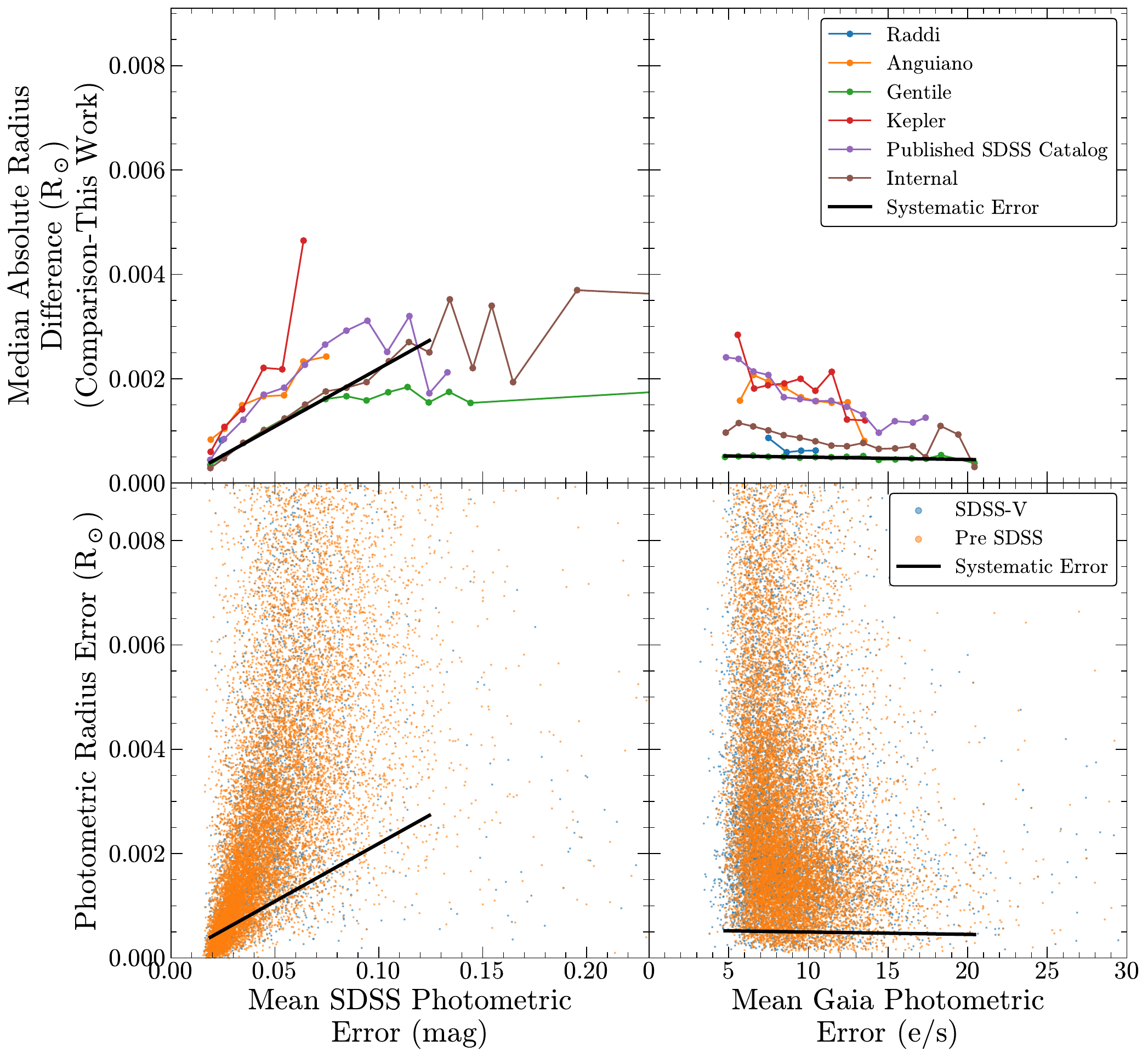}
\caption{The uncertainty on the radius as a function of photometric error. The top plot characterizes the systematic uncertainty in our SDSS (left) and Gaia (right) photometric radius fits by comparing our measurements to other data sets. The blue, orange, green, red, and purple data points show the median absolute difference in each photometric error bin when comparing our measured radii to the \citet{Raddi_2022}, \citet{Anguiano_2017}, \citet{Gentile_2021}, \citet{Kepler_2019}, and published SDSS WD catalogs, respectively. The brown data points show an internal comparison in which we compare our measured radii using SDSS and Gaia photometry. The bottom plot shows the measurement uncertainty returned by our fitting routine for the SDSS-V (blue) and the previous SDSS (orange) catalogs. The black curve in the left plots is the systematic uncertainty from a linear fit to the brown points, and the functional form of this curve is given in Eqn. \ref{eqn:rad_phot_err_sdss}. The black curve in the right plots is the systematic uncertainty from a linear fit to the green points, and the functional form of this curve is given in Eqn. \ref{eqn:rad_phot_err_gaia}. \label{fig:rad_err_photerr}}
\end{center}
\end{figure}

\begin{figure}[h!]
\begin{center}
\includegraphics[scale=0.45]{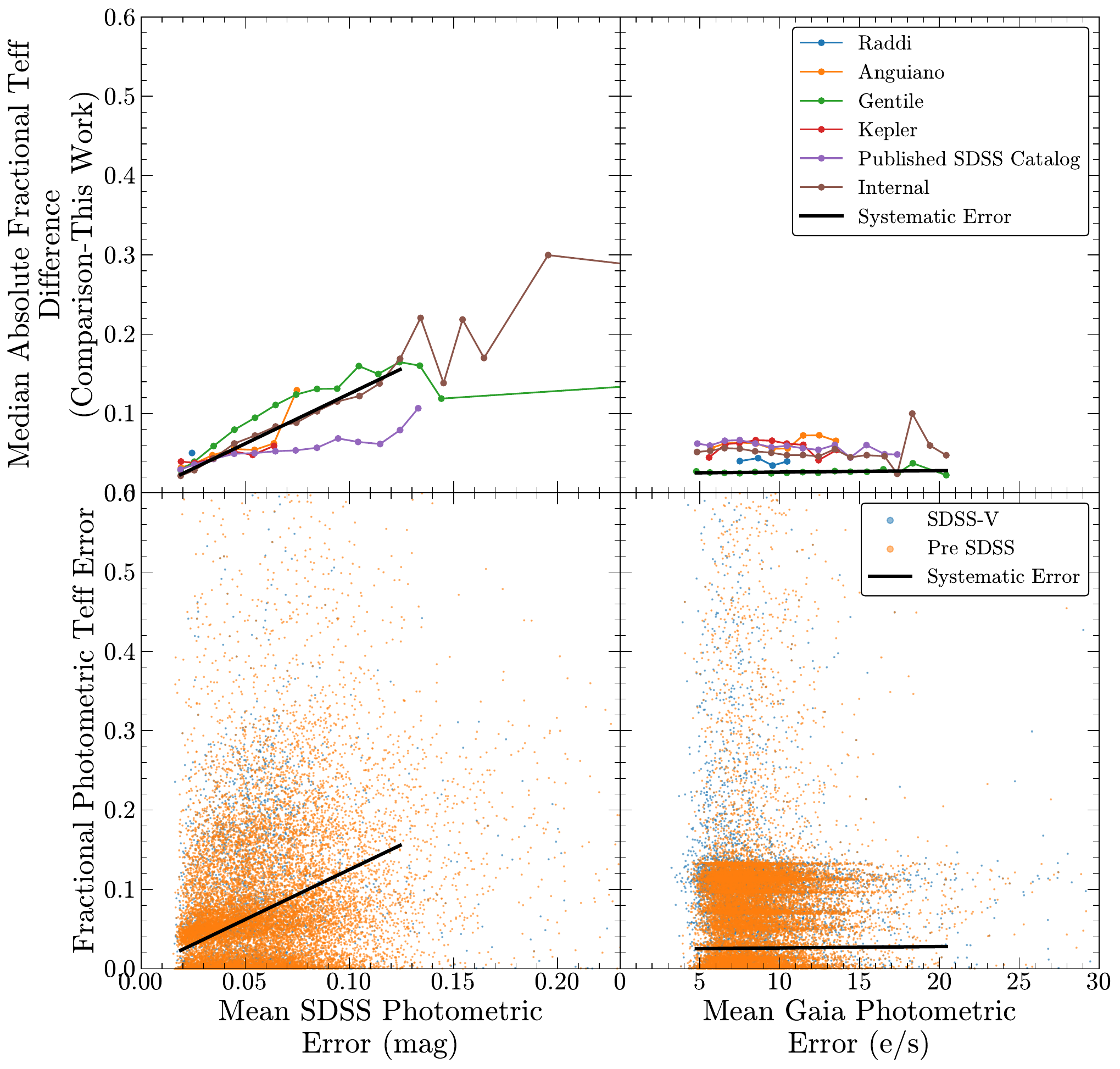}
\caption{Same as Fig. \ref{fig:rad_err_photerr}, but showing the fractional uncertainty on the measured effective temperature. The black curve in the left plots is the systematic uncertainty from a linear fit to the brown points, and the functional form of this curve is given in Eqn. \ref{eqn:teff_phot_err_sdss}. The black curve in the right plots is the systematic uncertainty from a linear fit to the green points, and the functional form of this curve is given in Eqn. \ref{eqn:teff_phot_err_gaia}. \label{fig:teff_err}\label{fig:teff_err_photerr}}
\end{center}
\end{figure}

\subsubsection{Validation and Systematic Uncertainties}

\indent We validate this fitting method by comparing our measured photometric radii and effective temperatures to the published values in the \citet{Raddi_2022}, \citet{Anguiano_2017}, \citet{Gentile_2021}, and \citet{Kepler_2019} data sets for SDSS-V objects and to \citet{Gentile_2021} and published SDSS WD catalog values for the previous SDSS objects. Although we remove any WDs flagged as potential binaries based on apparent radial velocity variation or Gaia RUWE, there are still binary contaminants in the catalog. So, we exclude WDs with large radii ($R>0.015~R_\odot$) that are likely in binaries \citep{Marsh_1995} from these comparisons. Such low mass WDs have low mass progenitors which would not have evolved into WDs over the lifetime of the universe, meaning they must have evolved in multiple star systems. 

\begin{figure}[h!]
\begin{center}
\includegraphics[scale=0.45]{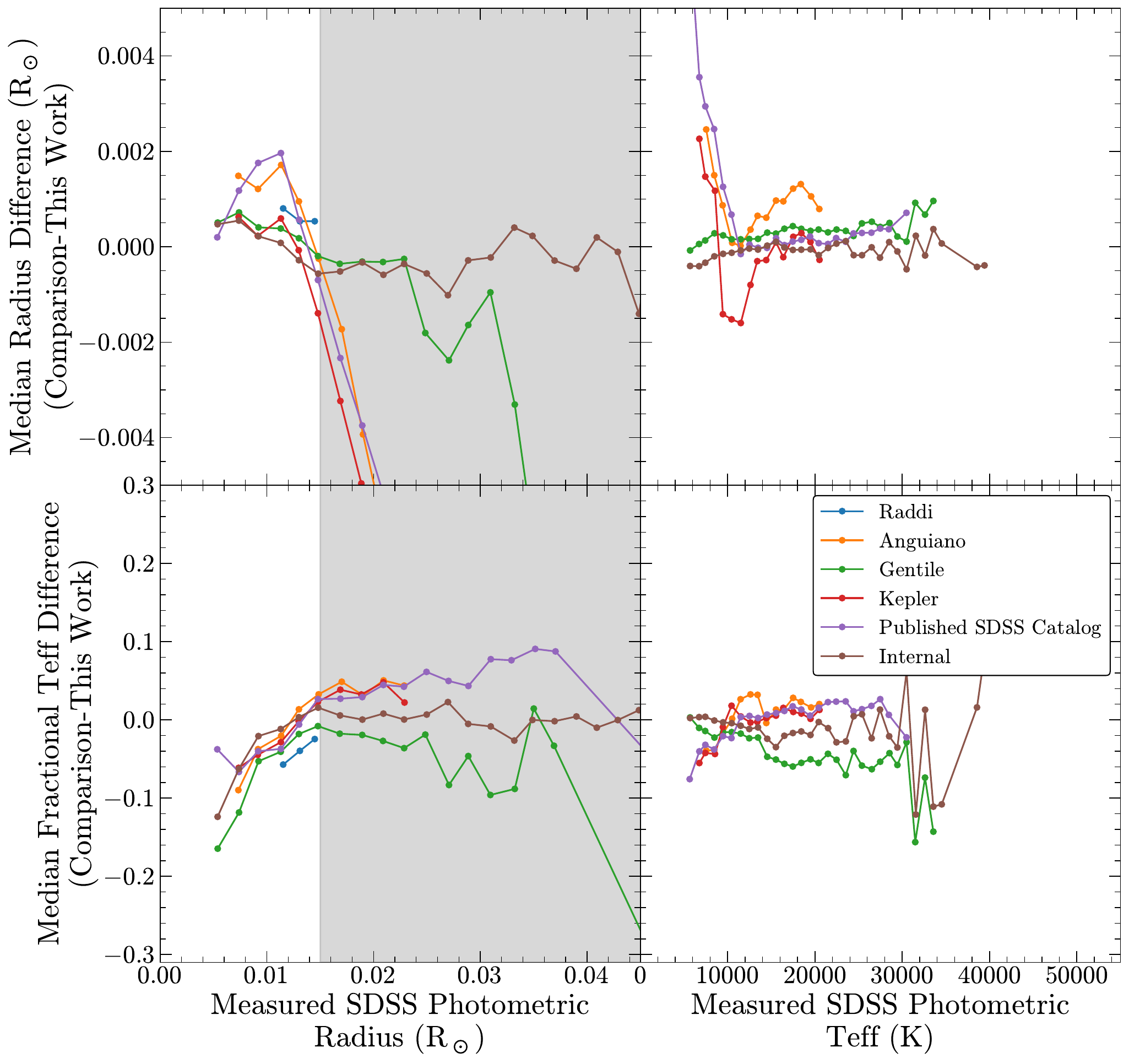}
\caption{The median difference between the radius (top) and median fractional difference between the effective temperature (bottom) measured using SDSS \textit{urz} photometry and from other datasets, as a function of our measured radius (left) and temperature (right). The blue, orange, green, red, and purple data points show the median absolute difference in each bin when comparing our measurements to the \citet{Raddi_2022}, \citet{Anguiano_2017}, \citet{Gentile_2021}, \citet{Kepler_2019}, and published SDSS WD catalogs, respectively. The brown data points show an internal comparison in which we compare our measured radii using SDSS and Gaia photometry. \label{fig:phot_err_sdss}}
\end{center}
\end{figure}

\begin{figure}[h!]
\begin{center}
\includegraphics[scale=0.45]{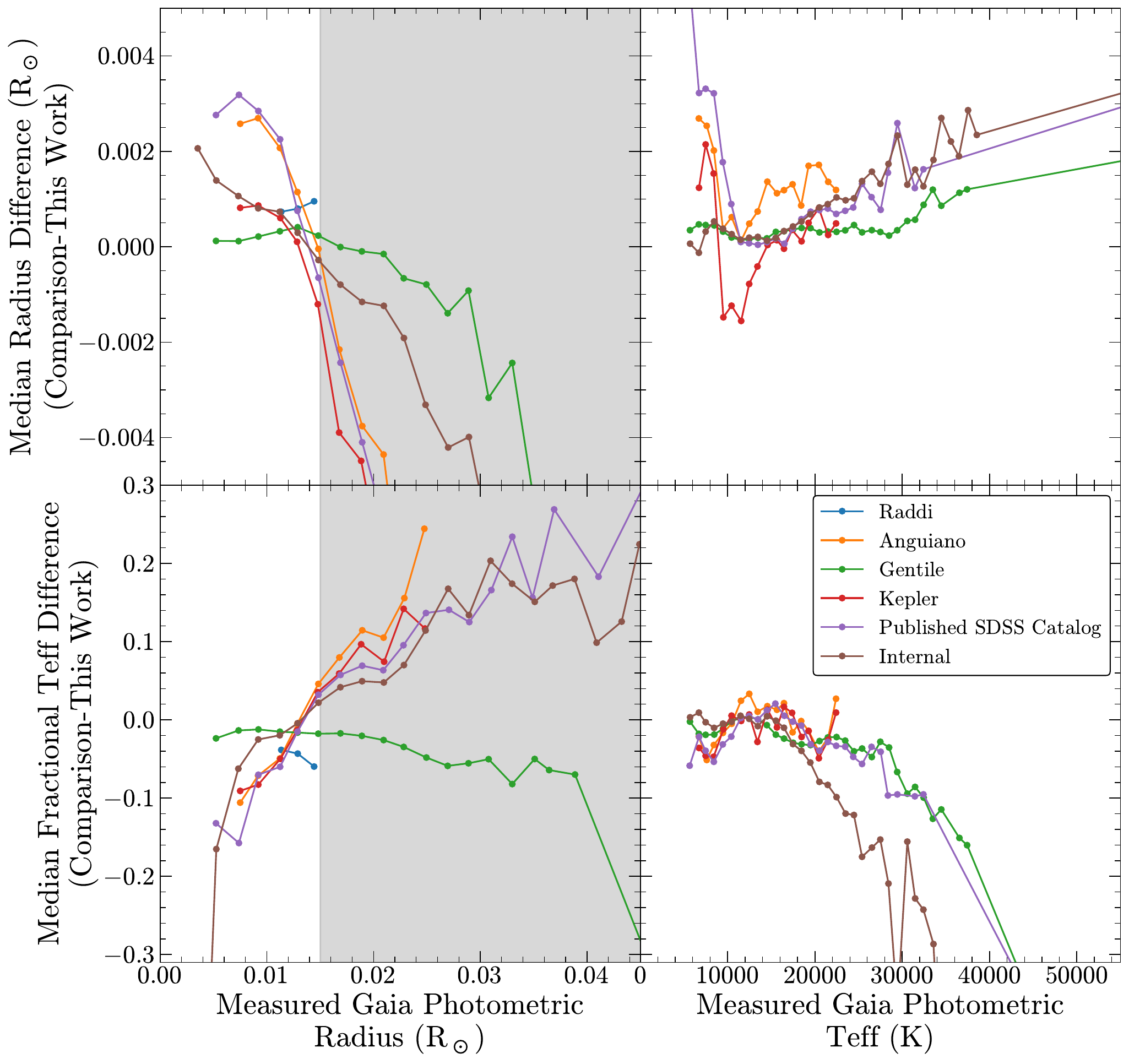}
\caption{Same as Fig. \ref{fig:phot_err_sdss}, but for measurements using Gaia $G_{\text{BP}}$ and $G_{\text{RP}}$ photometry. \label{fig:phot_err_gaia}}
\end{center}
\end{figure}

\indent We plot these comparisons in Figs. \ref{fig:rad_err_photerr}, \ref{fig:teff_err_photerr}, \ref{fig:phot_err_sdss}, and \ref{fig:phot_err_gaia}. Figs. \ref{fig:rad_err_photerr} and \ref{fig:teff_err_photerr} show the median absolute difference in the measured photometric radius and temperature, respectively, as a function of the SDSS or Gaia mean photometric error for all comparison data sets and for an internal comparison. For all comparisons, we include only bins with at least 20 entries. The mean SDSS photometric error is computed as the mean of the \textit{urz} photometric errors in AB magnitudes, and the mean Gaia photometric error is computed as the mean of the $G_{\text{BP}}$ and $G_{\text{RP}}$ flux errors in electrons per second. The internal comparison takes the difference between our photometric parameters measured with SDSS and Gaia photometry. The sign of this difference is chosen to be consistent with other comparisons, so if SDSS parameters are being subtracted from comparison set parameters then the difference is computed as our Gaia minus our SDSS measurements. 

\indent We use these comparison data sets to characterize the systematic uncertainty in measuring photometric parameters with this method as a function of photometric error. For the Gaia photometry, the best comparison data set is the \citet{Gentile_2021} catalog, which measures WD radii and effective temperatures using the same Gaia photometry employed in our measurements, although with different methods of determining distances and accounting for reddening as well as assuming a mass-radius relation instead of leaving temperature, radius, and surface gravity as independent parameters. The median absolute error between our measurements and those of \citet{Gentile_2021} is to within $0.0005$ $R_\odot$ ($4\%$) for photometric radii and to within $3\%$ for effective temperature across all Gaia photometric errors. This agreement is excellent and validates our fitting routine. To characterize the systematic uncertainty in our Gaia photometry fit routine, we perform a linear fit to the median absolute difference between our measurements and those of \citet{Gentile_2021}, the green data points from Figs. \ref{fig:rad_err_photerr} and \ref{fig:teff_err_photerr}, resulting in
\begin{align}
    \sigma_{\text{R, sys}}&=0.0005 R_\odot\label{eqn:rad_phot_err_gaia}\\
    \sigma_{\text{Teff, sys}}&=0.0002 \sigma_{\text{phot}}+0.024 \text{ (fractional)},\label{eqn:teff_phot_err_gaia}
\end{align}
where $\sigma_{\text{phot}}$ is the mean photometric error in the $G_{\text{BP}}$ and $G_{\text{RP}}$ bands in electrons per second.

\indent The best comparison set to characterize the systematic error on our fits using SDSS photometry is less clear. We include the comparisons to the radii from the \citet{Anguiano_2017}, \citet{Kepler_2019}, and published SDSS catalogs for completeness, but all of these radii were measured using spectroscopic surface gravities and the mass-radius relation, not from photometry. The agreement with these catalogs for radii measured with either SDSS or Gaia photometry is to within $\sim 0.001-0.002$ $R_\odot$. The precision of WD radii determinations is limited by the uncertainties in distance to about $0.001-0.002$ $R_\odot$ \citep{Chandra_2020_1}. We conclude this technique is in excellent agreement with the other catalogs despite the different measurement techniques. However, this agreement is significantly worse than the agreement between our Gaia fits and \citet{Gentile_2021} due to systematic differences between measurements from spectroscopy and photometry. This makes any of these sets poor choices to characterize the systematic error on our radius fits. Both the \citet{Raddi_2022} and \citet{Gentile_2021} parameters were measured using Gaia, not SDSS, photometry, meaning these comparisons suffer both from differences in fitting techniques and from systematic differences due to fitting different photometric data sets. Thus both the \citet{Raddi_2022} and \citet{Gentile_2021} data sets are poor choices to describe the systematic error of our radius fits from SDSS photometry. We conclude that the best comparison set to validate our SDSS photometric radii is our own set of Gaia photometric radii, which are at least measured using the same technique even if with a different photometric data set. This method also suffers from systematic differences between measurements with different instruments and likely overestimates the systematic error on our SDSS fits.

\indent For effective temperature, the agreement with the \citet{Anguiano_2017}, \citet{Kepler_2019}, and published SDSS spectroscopic catalogs for temperatures measured with either SDSS or Gaia photometry is to within $\sim 5-6\%$. This agreement is also excellent considering the very different measurement techniques employed by the comparison catalogs, but, like the radius comparison, is significantly worse than the agreement between our Gaia temperatures and those of \citet{Gentile_2021}. For consistency, we also characterize the systematic uncertainty in our SDSS effective temperature fits using our own measurements from Gaia photometry. Although, our photometric effective temperature measurements agree best with temperatures from published SDSS WD catalogs, and this also likely overestimates the error on the SDSS photometric temperature. The resulting linear fits to the median absolute difference between our SDSS and Gaia photometric fits, the brown data points from Figs. \ref{fig:rad_err_photerr} and \ref{fig:teff_err_photerr}, are
\begin{align}
    \sigma_{\text{R, sys}}&=0.0222\sigma_{\text{phot }} R_\odot\label{eqn:rad_phot_err_sdss}\\
    \sigma_{\text{Teff, sys}}&=1.3\sigma_{\text{phot}}-0.001 \text{ (fractional)},\label{eqn:teff_phot_err_sdss}
\end{align}
for SDSS photometry where $\sigma_{\text{phot}}$ is the mean photometric error in the \textit{urz} bands in magnitudes. 

\indent The overall SDSS median photometric error is $0.05$ mag, resulting in a systematic uncertainty of $6\%$ and $0.001$ $R_\odot$ for the measured temperature and radius, respectively. The overall Gaia median photometric error is $8$ electrons per second, resulting in a systematic uncertainty of $3\%$ and $0.0005$ $R_\odot$ for the measured temperature and radius, respectively. Although the systematic uncertainty on the Gaia parameters is smaller, we recommend the SDSS photometric parameters for use by the community because of the importance of including \textit{u}-band photometry in the fit. For all effective temperature and radius measurements, we add this systematic uncertainty in quadrature to the measurement uncertainty returned by our fitting routine to get the full error on these measurements. In Figs. \ref{fig:rad_err_photerr} and \ref{fig:teff_err_photerr}, we also plot the measurement uncertainty returned by the fit routine as a function of photometric error. This measurement uncertainty consists of the statistical uncertainty from the fit combined with the uncertainty from fixing the surface gravity in the fit (if applicable) combined with the uncertainty from the range of possible distances for the WD. We find that including the systematic uncertainty is important for fully characterizing the error on the photometric fits, since the fit uncertainties are often smaller than the systematic uncertainty. 

\indent In Figs. \ref{fig:rad_err_photerr} and \ref{fig:teff_err_photerr}, the error on the measured radius and effective temperature increases with increasing mean SDSS photometric error, but remains roughly constant with increasing Gaia photometric error. This is because these measurement errors are dominated by distance-dependent effects. For SDSS magnitudes, the error on the photometry increases with distance to the observed WD. In contrast, Gaia mean photometric errors remain roughly constant over the distances considered here. We consider systematic error as a function of photometric error, not distance, since the distance uncertainty has already been accounted for in the measurement error.

\indent In Fig. \ref{fig:teff_err_photerr}, there are horizontal streaks in the fractional errors on the Gaia effective temperatures as a function of Gaia photometric error. These streaks are due to the same structure appearing in plots of the measurement errors on the effective temperatures from the $\chi^2$ fit as a function of Gaia photometric error, and they are not due to structure in the actual measured effective temperatures. Each of these streaks corresponds to a different range of effective temperatures, with the highest streaks occurring at the lowest temperatures. This structure results from the fact that cooler WDs are dimmer, and thus harder to observe, than warmer WDs, and so cool WDs have larger measurement errors.

\indent We investigate any systematic trends with measured radii and temperatures in Figs. \ref{fig:phot_err_sdss} and \ref{fig:phot_err_gaia}, which show the median difference in measured radius or temperature as a function of the measured radius or temperature for the SDSS and Gaia photometric fits, respectively. For all comparisons, we include only bins with at least 20 entries. Across all fits, there is strong evidence that these photometric fits become unreliable when the WD is likely in a binary, $R>0.015~R_\odot$ \citep{Marsh_1995}. This motivates our $R<0.015~R_\odot$ quality cut employed in the creation of Figs. \ref{fig:rad_err_photerr} and \ref{fig:teff_err_photerr}, and the same cut is used when plotting median differences as a function of temperature in Figs. \ref{fig:phot_err_sdss} and \ref{fig:phot_err_gaia}. The data from Figs \ref{fig:phot_err_sdss} and \ref{fig:phot_err_gaia} is available online, and can be used to create transformation relations between the radii and effective temperatures measured in this work and those from the comparison catalogs.

We measure photometric radii with SDSS photometry, and find that $19\%$ of the 8545 and $36\%$ of the 19,257 WDs from the SDSS-V and previous SDSS catalogs, respectively, have radii $R>0.015~R_\odot$. When using Gaia photometry, $33\%$ and $37\%$ of WDs from the SDSS-V and previous SDSS catalogs, respectively, have radii larger than this threshold. These large percentages are due to data quality issues when measuring photometric parameters for distant WDs. When restricting the sample to WDs nearer than 300 pc, the percentages of WDs with radii $R>0.015~R_\odot$ are $6\%$ and $12\%$ when using SDSS photometry and are $12\%$ and $12\%$ when using Gaia photometry for the SDSS-V and previous SDSS catalogs, respectively. This is roughly consistent with the unresolved WD binary percentage in the literature, which varies from 1-10 \% \citep{Holberg_2009,Toonen_2017,Maoz_2018,Torres_2022}.

For measured SDSS temperatures and both  measured Gaia radii and temperatures, we find that our fits become more unreliable at small radii, when WDs are more difficult to observe because their small size produces less flux. Our SDSS radii still perform well at small sizes. The quality of our SDSS and Gaia radius fits shows no significant trends with measured temperature since the internal and \citet{Gentile_2021} agreement remains relatively constant. There is more disagreement with spectroscopic radii at low temperatures, but at low temperatures the strength of the Balmer lines weakens making spectroscopic fits more difficult. At high effective temperatures, our SDSS measured temperatures differ from those measured using Gaia data, both our own and the \citet{Gentile_2021} measurements. In the same range, our SDSS temperatures agree well with spectroscopic effective temperature measurements from published SDSS WD catalogs. This is likely because at these temperatures the WD spectral energy distribution becomes bluer, making the \textit{u}-band information provided by SDSS important while Gaia only has access to the red tail of the WD spectrum. \citet{Bergeron_2019} comprehensively discuss WD photometric measurements with different data sets.

\subsubsection{Comparison of Spectroscopic and Photometric Temperatures}\label{sec:spec_phot_comp}

\indent Spectroscopy and photometry provide two independent measurements of the effective temperature of the WD. We compare the photometric temperature to the spectroscopic temperature measured from the coadded spectrum of each WD for coadded spectra with a SNR$>20$ and mean SDSS or Gaia photometric errors $<0.05$ mag or $<10$ electrons/second, respectively. We find agreement to within $800-900$ K or $5-6$\%. However, spectroscopic temperature fits occasionally ($\sim2.5$\% of fits) result in anomalously hot temperatures relative to the photometric results. This may be due to a degeneracy between hot and cold solutions in fitting spectroscopic parameters \citep{Chandra_2020_1}. Because of this, we recommend that catalog users utilize our photometric temperatures rather than our spectroscopic temperatures.

\section{Other Parameters Included in Our Catalog} \label{sec:other}

\subsection{WD Masses} \label{sec:mass}

\begin{figure}[hb!]
\begin{center}
\includegraphics[scale=0.5]{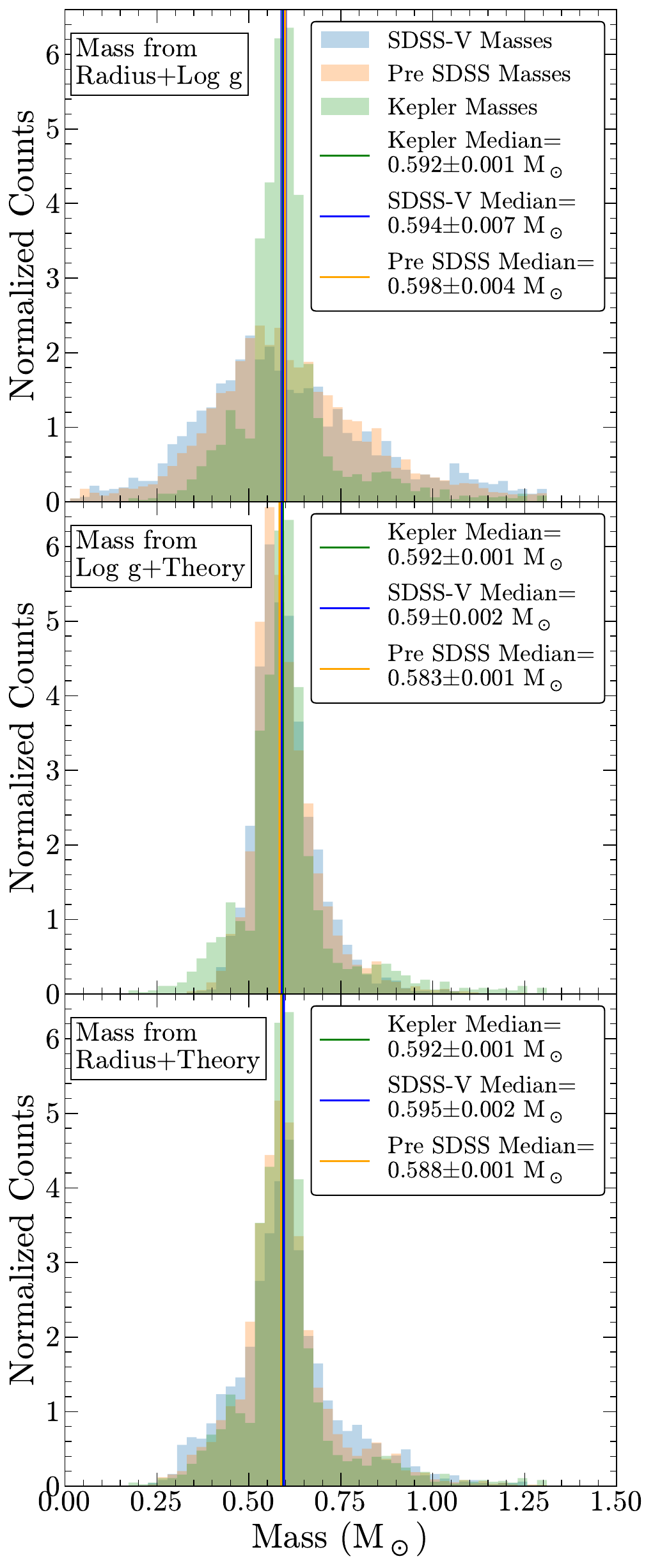}
\caption{Our mass distribution of DA WDs, measured with various methods. From top to bottom, we compare the \citet{Kepler_2019} mass distribution (green) to our SDSS-V (blue) and previous SDSS (orange) catalog mass distributions with masses measured from photometric radius combined with spectroscopic surface gravity, surface gravity combined with theoretical models, and radius combined with theoretical models. The vertical lines are the median masses of each sample. \label{fig:mass_dists}}
\end{center}
\end{figure}

\indent We include three different methods of determining individual WD masses in our catalog, based on measurements from coadded spectra. In the first method, we combine the measured spectroscopic surface gravity with the measured photometric radius to yield the mass for each object. This method has the benefit of being independent of an assumed WD mass-radius relation, but the use of these two measured parameters dramatically increases the noise in these mass measurements. In the remaining two methods, we combine either the measured surface gravity or the measured radius with the measured spectroscopic or photometric temperature, respectively, and theoretical models to obtain the mass of the WD. For these theoretical models, we utilize the La Plata models\footnote{\url {http://evolgroup.fcaglp.unlp.edu.ar/TRACKS/newtables.html}}, which contain tables of DA WD masses and radii as a function of effective temperature and surface gravity. For low-mass helium core WDs, intermediate-mass carbon-oxygen core WDs, and high-mass oxygen-neon core WDs, these models use the results of \citet{Althaus_2013}, \citet{Camisassa_2016}, and \citet{Camisassa_2019}, respectively. In these models, the thickness of the WD hydrogen envelope varies with the mass of the object, with thinner envelopes at lower masses. Because the determined mass of the WD is less sensitive to the measured temperature than to the measured radius or surface gravity, these mass measurements are far less noisy.

\indent In Fig. \ref{fig:mass_dists}, we plot our measured masses for the SDSS-V and previous SDSS catalogs for coadded spectra against the measured SDSS Data Release 14 DA WD masses from \citet{Kepler_2019}. \citet{Kepler_2019} produced their mass distribution by combining spectroscopic surface gravity measurements with a theoretical WD mass-surface gravity relation. In all cases, we include only WDs for which the spectra used have SNR$>20$. We exclude objects flagged as binaries from our measured mass distributions. We measure the median mass for each sample, and derive the uncertainty on the median by bootstrapping each sample 1000 times, re-calculating the median for each bootstrap, and taking the standard deviation of the median across all bootstraps. The median masses of our SDSS-V and previous SDSS catalogs fluctuate depending on the measurement method, but are mostly in agreement with the median mass of the \citet{Kepler_2019} catalog (within $2\sigma$). The exceptions are our mass distributions from radii or surface gravities and theory for the previous SDSS catalog, which have median masses much less ($\gtrsim3\sigma$) than the \citet{Kepler_2019} median mass. Our measured mass distributions derived from both measured radii and measured surface gravities are broader than the \citet{Kepler_2019} distribution. This is because both the measured radius and measured surface gravity of each star are noisy, resulting in more uncertain mass measurements and corresponding to a less sharply peaked distribution. Our mass distributions derived from surface gravity and the La Plata theoretical models are narrower than the \citet{Kepler_2019} distribution, and our mass distributions derived from the measured radius and the La Plata theoretical models are broader than the \citet{Kepler_2019} distribution. Both distributions are generally in agreement with the \citet{Kepler_2019} distribution, and thus either masses from photometric radii or spectroscopic surface gravities can be used by the community.

\subsection{Stellar Populations} \label{sec:stellarpop}

\indent We include a flag in the catalog to indicate which stars are likely part of the Milky Way thin disk population, part of the halo population, and part of the thick disk population or have an uncertain population designation. Stars which are part of the thin disk population have relatively small speeds and stars which are part of the halo population have large speeds relative to the local standard of rest \citep[LSR, ][]{Binney_1987}. We use Gaia proper motion information to obtain a transverse speed for each WD relative to the LSR. Using the measured photometric radius and effective temperature for each WD combined with the La Plata models, we obtain a theoretical gravitational redshift for each star. Subtracting this gravitational redshift from the apparent radial velocity gives the intrinsic radial velocity for the star, which we convert to the LSR frame. We combine the transverse and LSR-corrected intrinsic radial velocity to measure the total speed of the WD relative to the LSR. We use a Toomre diagram to flag any WD for which this total speed is $<100$ km/s as a thin disk object \citep{Raddi_2022} and any WD for which this total speed is $>220$ km/s as a halo object \citep{Du_2018}. Of the 26,041 unique WDs contained in this catalog, 20,453 objects likely belong to the Milky Way thin disk, 476 likely belong to the Milky Way halo, and 5,112 belong to the Milky Way thick disk or have uncertain designations.

\begin{figure}[h!]
\begin{center}
\includegraphics[scale=0.45]{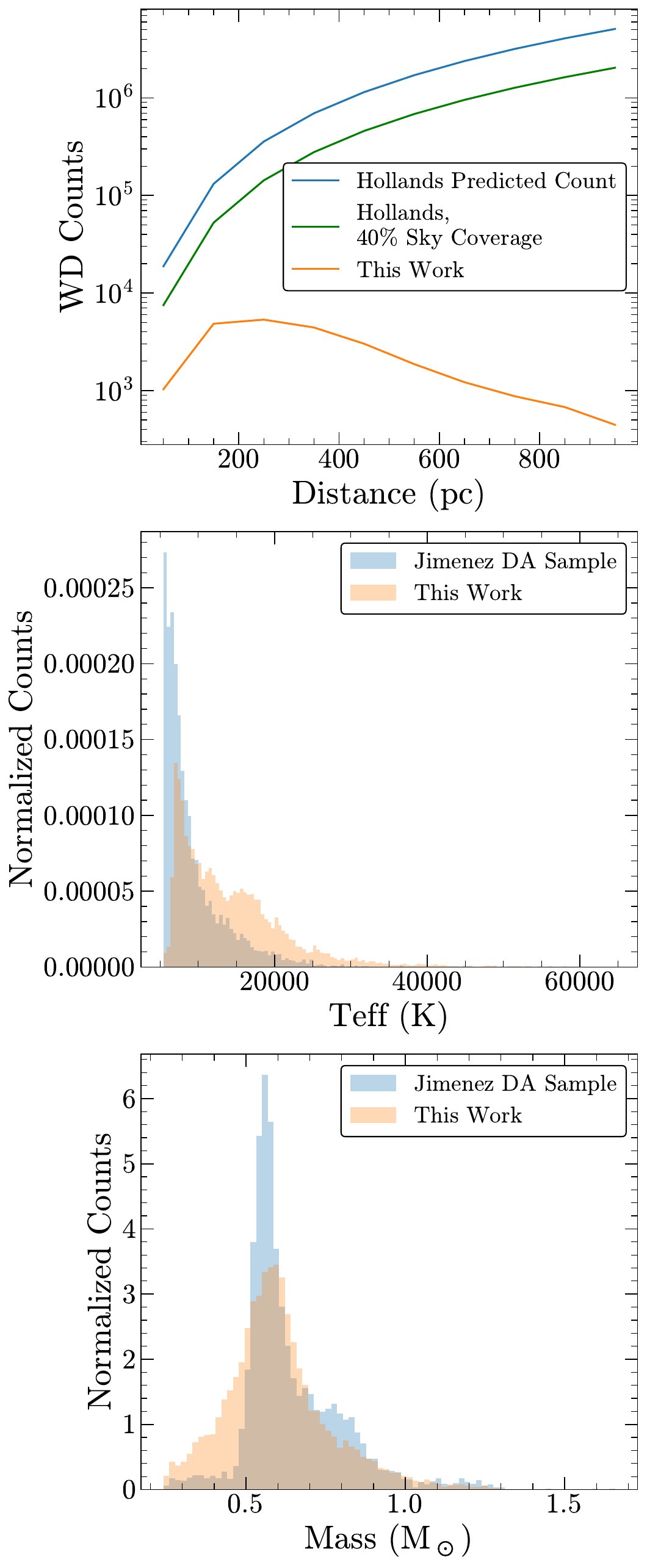}
\caption{Completeness as a function of WD distance (top), effective temperature (middle), and mass (lower). To characterize the completeness of our catalog (orange) we compare to the \citet{Hollands_2018} (top, blue) or \citet{Jimenez_2023} (middle and bottom, blue) highly complete samples. In the top panel, we also include the expected distance completeness from \citet{Hollands_2018}, adjusted for $40\%$ sky coverage (green).\label{fig:complete}}
\end{center}
\end{figure}

\subsection{Completeness Estimate} \label{sec:complete}

\indent Spectroscopic samples of WDs, such as this catalog, are well-known to be severely incomplete and biased due to choices in observing strategy \citep{Gentile_2021}. To investigate these effects, we characterize the completeness of our sample as a function of WD distance, mass, and effective temperature in Fig. \ref{fig:complete}. To measure completeness as a function of distance, we use the results of \citet{Hollands_2018}, who used a highly complete sample of WDs within 20 pc of the Sun to estimate a space-density of WDs of $4.49\times10^{-3}$ $\text{pc}^{-3}$. In the top panel of Fig. \ref{fig:complete}, we use this density of WDs to measure the completeness of our sample as a function of distance. We bin our catalog in distance intervals of 100 pc, and compare the number of WDs in that bin to the expected number given the volume of that spherical shell. Additionally, we calculate the sky coverage of our catalog by dividing the sky into a grid $3\text{ deg}^2$ squares, and flagging any square containing at least one WD from our catalog. We apply this method to all DA WDs in our catalog within 1000 pc, and find that our catalog covers $\sim40\%$ of the sky. In Fig. \ref{fig:complete}, we also plot the \citet{Hollands_2018} space-density adjusted to account for the limited sky coverage of our sample. As expected, we find that our catalog is severely incomplete. When adjusting for $40\%$ sky coverage, our catalog contains only $13.7\%$ of the expected number of WDs within 100 pc, and this percentage decreases with distance. Within the 900 to 1000 pc bin, our catalog contains only $0.02\%$ of the expected number of WDs. Even accounting for the fact that the \citet{Hollands_2018} density includes all spectroscopic types of WDs, not solely DA WDs, our catalog is severely incomplete.

\indent To characterize completeness as a function of mass and effective temperature, we compare our catalog to the highly complete sample of \citet{Jimenez_2023}. The \citet{Jimenez_2023} sample is $\sim 70\%$ complete relative to the expected WD density from \citet{Hollands_2018}, and most of the missing sources are on the reddest end of the WD sequence. \citet{Jimenez_2023} were able to classify $99\%$ of objects in their catalog with effective temperatures $>5,500$ K as DA or non-DA WDs. In Fig. \ref{fig:complete}, we compare our measured DA WD photometric effective temperatures and masses to the distributions of these parameters for the DA WDs in the \citet{Jimenez_2023} catalog. We use our masses measured from photometric radii and the La Plata models. We find that our sample contains proportionally more high temperature and low mass WDs compared to the \citet{Jimenez_2023} sample. This is in line with our expectations as hotter and less massive WDs are larger and brighter, and thus easier to observe. Because this catalog is incomplete and biased, catalog users should employ caution if using this catalog for certain analyses, such as creating a WD luminosity function to determine the age of the Galactic disk.

\subsection{Temperature Dependence Detection Parameters} \label{sec:tempdep}

\indent In our companion paper \citep{Crumpler_2024}, we use this catalog to detect the temperature-dependence of the WD mass-radius relation. We include a flag in our published catalog to indicate whether each SDSS-V or previous SDSS observation was used in the \citet{Crumpler_2024} detection. For WDs used in the temperature-dependence detection, we provide apparent radial velocities which have been corrected to the LSR, the locally co-moving frame of stars within the solar neighborhood, in our published catalogs. Only objects nearer than $500$ pc were used in the temperature-dependence detection, as the LSR-corrected apparent radial velocities are only valid for nearby objects. Additionally, because WDs are an older stellar population relative to main sequence stars, WDs experience an asymmetric drift relative to the LSR \citep{Binney_1987}. This asymmetric drift can be connected to the Galactic velocity dispersion of the stellar population \citep{Schonrich_2010}, which varies depending on the selection cuts applied to our catalogs. In our published catalogs, we include the asymmetric drift-corrected apparent radial velocities for objects used in the temperature-dependence detection. For different subsets of WDs, these corrections must be re-calculated. Catalog users can access the public source code used to correct for asymmetric drift in our measurements, and apply those methods to their own subsets of these catalogs.

\section{SDSS-V and Previous SDSS Catalog Parameter Distributions} \label{sec:SDSS_diffs}

\begin{figure*}[h!]
\begin{center}
\includegraphics[scale=0.235]{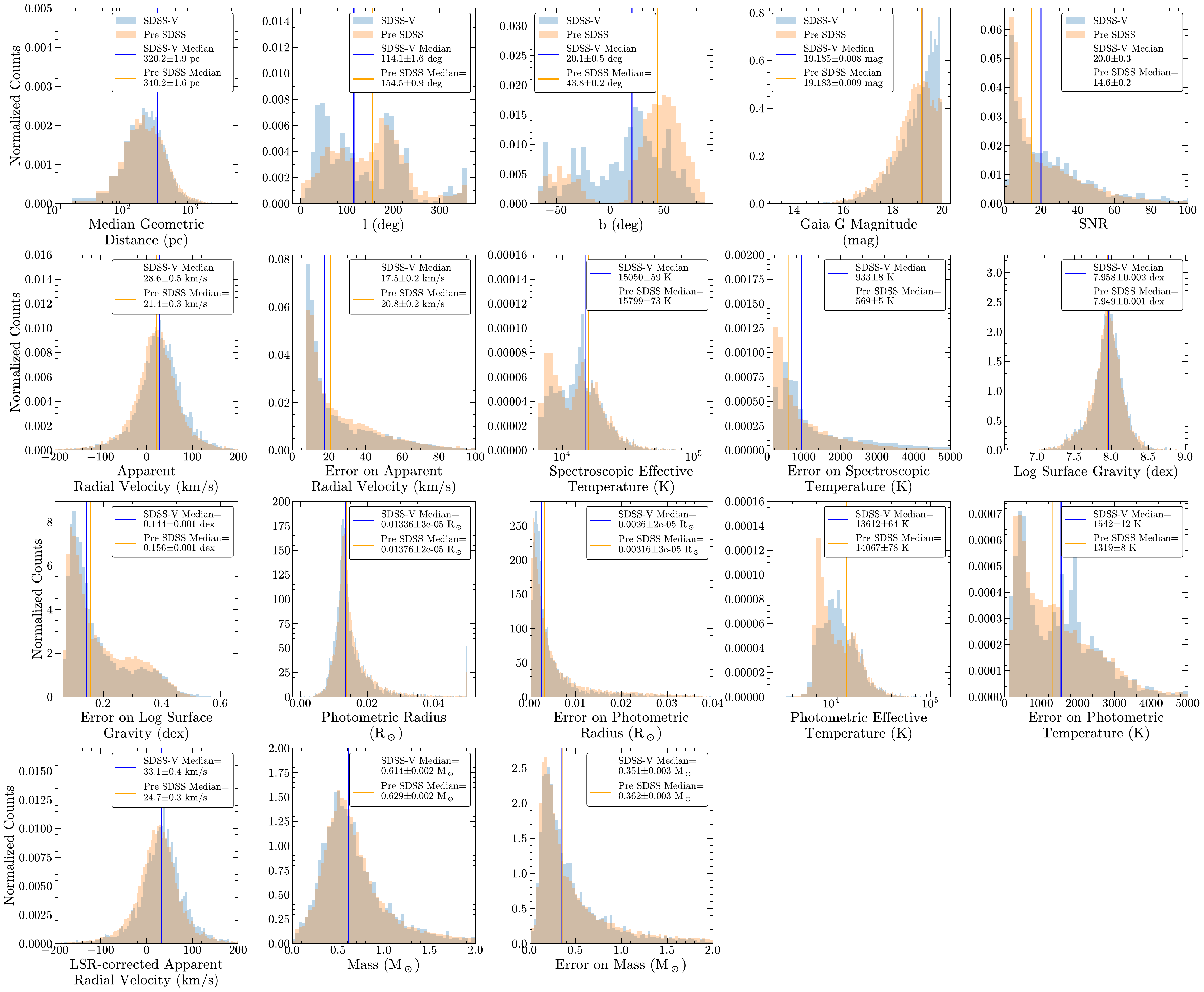}
\caption{Distributions of all relevant physical parameters for the SDSS-V (blue) and previous SDSS (orange) catalogs. From left to right and top to bottom, we show the distributions of the WD median geometric distances, Galactic $l$ coordinates, Galactic $b$ coordinates, Gaia $G$-band magnitudes, coadded spectrum SNRs, coadded spectrum apparent radial velocities, full errors on the apparent radial velocities, coadded spectrum effective temperatures, full errors on the spectroscopic temperatures, coadded spectrum $\log$ surface gravities, full errors on the $\log$ surface gravities, photometric radii, full errors on the radii, photometric effective temperatures, full errors on the photometric temperatures, LSR-corrected apparent radial velocities, masses, and full errors on the masses. The vertical lines are the medians of each parameter for each sample. \label{fig:SDSS_dists}}
\end{center}
\end{figure*}

\indent In this section, we investigate the systematic differences in the measured parameters of our SDSS-V and previous SDSS catalogs. Some of these differences are discussed briefly in \citet{Crumpler_2024}. Namely, when using a subset of 2887 SDSS-V and 7242 previous SDSS WDs passing data quality cuts, \citet{Crumpler_2024} finds differences in the mean gravitational redshifts of the SDSS-V and previous SDSS samples, which are $31.9$ and $27.0$ km/s, respectively. They also find differences in the velocity dispersions of the SDSS-V and previous SDSS samples, which are $37$ and $34$ km/s, respectively. These differences are of moderate size relative to the $\sim 2$ km/s dispersion found in comparing measured apparent radial velocity from coadded spectra to the weighted mean of apparent radial velocities from individual spectra. Systematic differences between our SDSS-V and previous SDSS catalogs can appear in two main ways. In the first, distributions of the same measured parameter from the SDSS-V and previous SDSS catalogs can be different due to different populations of WDs being observed in either data set. In the second, for the same WD observed in both catalogs, measurements of the same parameter from SDSS-V and previous SDSS spectra can also be different due to differences in survey hardware and reduction pipeline choices. 

\subsection{Differences from Observed White Dwarf Populations}

\indent When investigating parameter distributions across the full SDSS-V and previous SDSS catalogs, we find systematic differences indicative of observing different populations of WDs. These discrepancies are the result of key differences in survey strategy in previous generations of SDSS and SDSS-V. In previous generations, nearly all WDs were identified serendipitously, having initially been targeted as potential quasars. So, these WDs are biased to be located out of the plane of the Milky Way and to have particular (\textit{u}-\textit{g},\textit{g}-\textit{r}) colors. This bias results in different spatial distributions of WDs in the previous SDSS and SDSS-V catalogs. Additionally, the color selection cuts used to identify WDs in previous generations enforce a temperature selection bias not present in the SDSS-V catalog. Cool WDs are underrepresented in previous generations of SDSS, so SDSS-V specifically targeted some, cooler WDs. Such differences in survey strategies result in differences in overall parameter distributions given that the WDs observed in either data set have distinct characteristics. 

\indent In Fig. \ref{fig:SDSS_dists}, we plot distributions of all relevant physical parameters for the full SDSS-V and previous SDSS catalogs with no data quality cuts. These include the spatial locations of the WDs as indicated by their median geometric distances and Galactic ($l,b$) coordinates, the brightness of the WDs from their Gaia $G$-band magnitudes, the quality of the spectra in each catalog set by the coadded spectrum SNR, the spectroscopic parameters and full errors on those parameters measured from the coadded spectra including the apparent radial velocity, surface gravity, and spectroscopic effective temperature, the photometric parameters and full errors on those parameters such as the radius and photometric effective temperature, the LSR-corrected apparent radial velocities, and the masses and full errors on the masses measured from the combined spectroscopic surface gravity and photometric radius for each object. All histograms are plotted in terms of normalized counts to account for the fact that the SDSS-V data set is much smaller than the previous SDSS catalog. For each parameter, we measure the median for each catalog, and derive the uncertainty of the median by bootstrapping each sample 1000 times, re-calculating the median for each bootstrap, and taking the standard deviation of the median across all bootstraps.

\indent Fig. \ref{fig:SDSS_dists} shows a clear difference in survey strategy between previous SDSS and SDSS-V WDs. Proportionally, SDSS-V contains more nearby WDs than earlier generations. The SDSS-V survey also contains far more WDs in the plane ($b\sim0$) compared to previous generations of SDSS, in which most WDs were initially targeted as quasars. This has important implications for photometric measurements, as the extra dust in the plane increases the dependence on the accuracy of our reddening corrections. The Gaia $G$-band magnitude distribution for SDSS-V peaks at higher magnitudes than for previous generations of SDSS, potentially indicative of SDSS-V targeting cooler WDs which are also dimmer. The presence of more faint WDs could also bias the SDSS-V sample towards more massive WDs, which are smaller and thus fainter than their less massive counterparts. Proportionally, the coadded spectrum SNR distribution of SDSS-V has more objects with $15<$SNR$<100$ than previous generations of SDSS, for which the peak at SNR$\sim5$ is sharper.

\indent Some of the spectroscopic parameter distributions in Fig. \ref{fig:SDSS_dists} are different between previous generations of SDSS and SDSS-V. We find that the SDSS-V apparent radial velocities are systematically larger than the previous SDSS values, even when correcting to the LSR. This is in line with the results of \citet{Crumpler_2024}. Since the LSR and asymmetric-drift-corrected apparent radial velocities can be averaged to yield the gravitational redshifts for the whole sample, this suggests that the masses from apparent radial velocities should be systematically larger (larger gravitational redshift) in SDSS-V. In Fig. 2 of \citet{Crumpler_2024}, we indeed find that SDSS-V WDs are more massive than previous SDSS WDs. The surface gravity distribution of the SDSS-V WDs is comparable to that of previous SDSS WDs, but the SDSS-V distribution has slightly larger surface gravities which would indicate the presence of more massive WDs. The spectroscopic effective temperature distributions are also different across the catalogs, with the previous SDSS WDs showing a stronger bimodal distribution in temperature than the SDSS-V objects, potentially due to color selection cuts and other targeting effects.

\indent For the photometric parameter distributions of Fig. \ref{fig:SDSS_dists}, we also find some systematic differences. The photometric radii distributions are largely in agreement, but with slight evidence that the SDSS-V photometric radii are smaller. This would suggest that SDSS-V contains more massive WDs than previous generations. This is in agreement with the systematic offset in apparent radial velocity between the two catalogs. As with the spectroscopic temperature, the previous SDSS photometric temperatures show a stronger bimodal distribution than the SDSS-V measurements. The mass distributions for both samples, measured by combining photometric radii with spectroscopic surface gravities, are comparable. However, the trend with mass reverses and the previous SDSS distribution is slightly shifted towards larger masses. In Fig. \ref{fig:SDSS_dists} the mass distributions peak at $\sim0.62~M_\odot$ while in Fig. \ref{fig:mass_dists} the mass distributions peak at $\sim0.59~M_\odot$. This discrepancy is due to differences in data quality cuts. A SNR$>20$ cut is implemented in Fig. \ref{fig:mass_dists}, and there are no data quality cuts in Fig. \ref{fig:SDSS_dists}.

\subsection{Differences When Observing the Same White Dwarf}

\begin{figure}[h!]
\begin{center}
\includegraphics[scale=0.4]{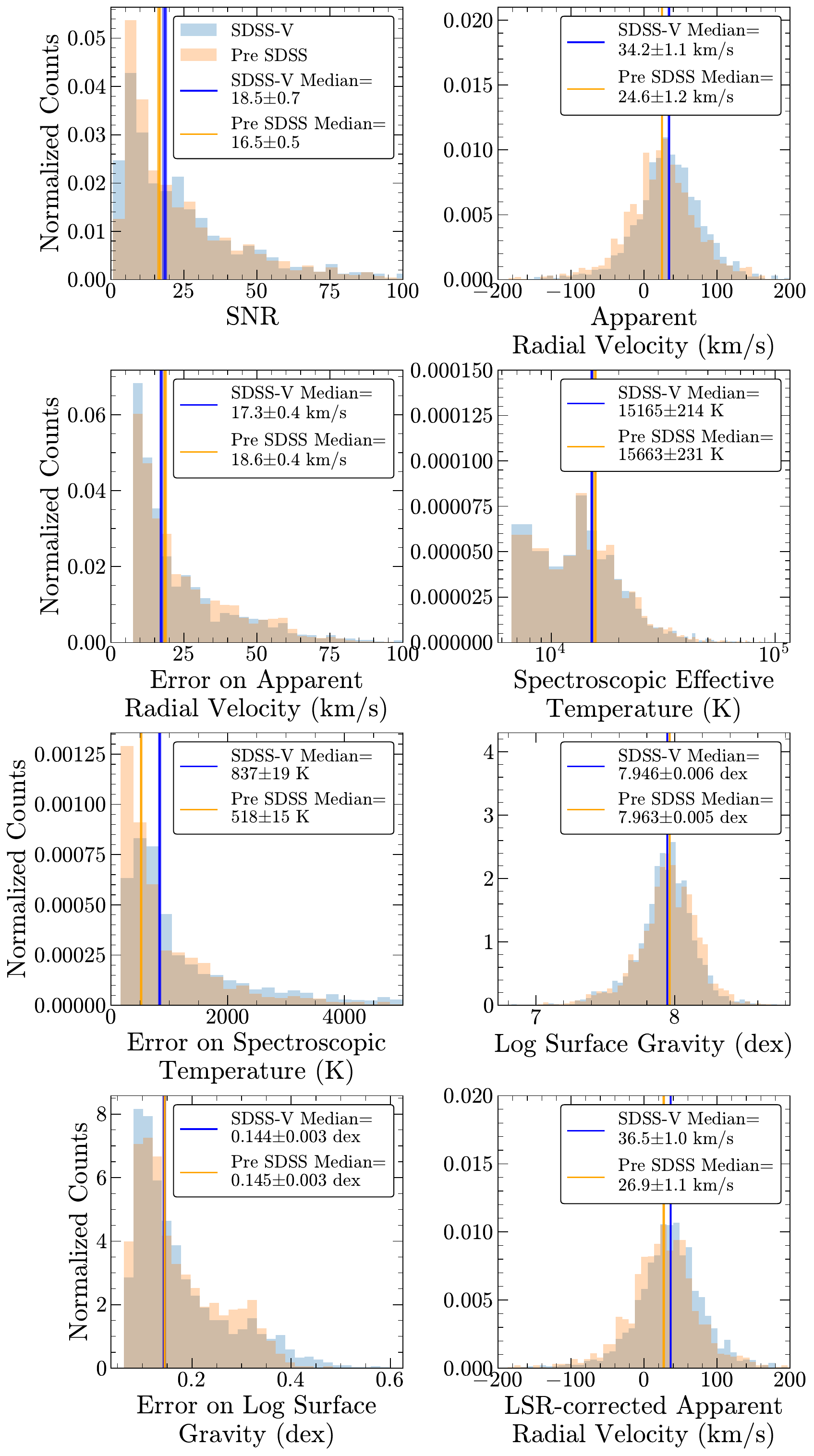}
\caption{Distributions of all spectroscopic parameters for the 1761 WDs contained in both the SDSS-V (blue) and previous SDSS (orange) catalogs. From left to right and top to bottom, we show the distributions of the WD coadded spectrum SNRs, coadded spectrum apparent radial velocities, full errors on the apparent radial velocities, coadded spectrum effective temperatures, full errors on the spectroscopic temperatures, coadded spectrum $\log$ surface gravities, full errors on the $\log$ surface gravities, and LSR-corrected apparent radial velocities. The vertical lines are the medians of each parameter for each sample. \label{fig:SDSS_dists_overlap}}
\end{center}
\end{figure}

\indent We repeat the above analysis for the subset of 1761 WDs contained both in the SDSS-V and previous SDSS catalogs. We investigate only the systematic differences in spectroscopic parameters, since all other parameters are the same between the SDSS-V and previous SDSS catalogs. In Fig. \ref{fig:SDSS_dists_overlap}, we plot count-density histograms of the coadded spectrum SNR, apparent radial velocity, surface gravity, and spectroscopic effective temperature, as well as the errors on these quantities. For each parameter, we measure the median for each catalog, and derive the uncertainty on the median by bootstrapping each sample 1000 times, re-calculating the median for each bootstrap, and taking the standard deviation of the median across all bootstraps. The SNR and effective temperature distributions of the same WDs observed in both SDSS-V and previous SDSS generations are comparable. However, there are unexpected offsets in the apparent radial velocity and the spectroscopic surface gravity distributions. For the surface gravity offset, on average the same WD observed in both in SDSS-V and previous SDSS generations has a measured surface gravity that is $0.015$ dex smaller when using SDSS-V data instead of previous SDSS data. When considering the median difference instead, this offset is $0.017$ dex. Similarly, on average, the same WD observed in both in SDSS-V and in previous SDSS generations has a measured apparent radial velocity that is $11.5$ km/s larger when using SDSS-V data instead of previous SDSS data. When considering the median difference instead, this offset is $9.6$ km/s. We find that the velocity dispersions of WDs contained in both the SDSS-V and previous SDSS catalogs are $66.0$ and $74.6$ km/s when using SDSS-V and previous SDSS spectra, respectively. These differences are large relative to the $\sim 2$ km/s dispersion found in comparing measured apparent radial velocity from coadded spectra to the weighted mean of apparent radial velocities from individual spectra.

\indent We explore the distribution of the difference in SDSS-V and previous SDSS apparent radial velocities and surface gravities for the same WD and find no strong skew accounting for these offsets, the means of the distributions are simply shifted to indicate higher apparent radial velocities and lower surface gravities in SDSS-V. We also investigate these offsets as a function of SNR, and find that the scatter of these differences increases with decreasing SNR, as expected. But, there is no correlation with SNR explaining the systematic trends. We explore the difference in apparent radial velocity as a function of measured SDSS-V or previous SDSS apparent radial velocity, and find positive and negative correlations, respectively. Similarly, we investigate the difference in surface gravity as a function of measured SDSS-V or previous SDSS surface gravity, and also find positive and negative correlations, respectively. 

\indent We re-measure all apparent radial velocities using only the $H\alpha$ and $H\beta$ Balmer series lines. These apparent radial velocities are not as accurate as those measured with the first four lines, since the median absolute agreement with all comparison data sets worsens by $1-2$ km/s when fitting fewer lines. However, we find that the discrepancy between SDSS-V and previous SDSS apparent radial velocities decreases from $11.5$ to $4.7$ km/s when only using the first two Balmer series lines. Such a difference could be explained by new telescope hardware or changes to the reduction pipeline. There has been work in SDSS-V to improve the stability of apparent radial velocity measurements by changing the treatment of skylines at the blue end of the BOSS spectrograph. These new methods change the wavelength solution for BOSS spectra and could cause this discrepancy. Members of the SDSS collaboration are currently working to uncover the effects of the new wavelength solution at the blue end of the BOSS spectrograph on WD apparent radial velocities and other spectroscopic analyses (Arseneau et al., in prep).

\section{Discussion and Conclusions} \label{sec:conc}

\indent In this paper, we have constructed catalogs of measured apparent radial velocities, surface gravities, effective temperatures, and radii for all DA WDs observed in previous generations of SDSS and in the 19th Data Release of SDSS-V. We have validated our measurements against published catalogs of WD physical parameters, and determined that our results are in agreement with these catalogs. For spectroscopic apparent radial velocities, surface gravities, and effective temperatures, the most valuable comparison comes from investigating the agreement of our results with previously published SDSS WD catalogs, which use the same spectra as our previous SDSS catalog. In this direct comparison, we find agreement to within 7.5 km/s (9 km/s), 0.060 dex (0.095 dex), and 2.4\% (2.5\%) for our apparent radial velocities, surface gravities, and effective temperatures measured from coadded spectra with SNR $>50$ (SNR $>20$). For photometric radii and effective temperatures, the most direct comparison results from comparing our measurements using Gaia photometry to those of \citet{Gentile_2021}, which also used fits to Gaia photometry. In this comparison, our results agree with the published values to within $0.0005$ $R_\odot$ and $3\%$ for radius and effective temperature measurements, respectively. However, we recommend that catalog users prioritize photometric fit parameters derived with SDSS photometry, as the SDSS \textit{u}-band constraint becomes increasingly important at high effective temperatures. For each WD, there are multiple measurements of the same parameter with different methods. We flag each column containing the best measurement for each unique WD in Appendices \ref{sec:app1} and \ref{sec:app2}, and suggest that members of the community use these values. We make this catalog\footnote{\url{https://www.sdss.org/dr19/data_access/value-added-catalogs/?vac_id=10008}} of all SDSS DA WDs and the code\footnote{\url {https://github.com/nicolecrumpler0230/WDparams}} used to create it publicly available so the results are reproducible and can be used by the community. This is the largest catalog of its kind currently available.

\indent Using our catalog, we obtain three different measurements of each DA WD mass. In the first method, we combine our measured surface gravity and radius to obtain the mass independent of an assumed mass-radius relation. Comparing the distribution of these masses with that of \citet{Kepler_2019}, we find that the sharp peak of the WD mass distribution is flattened by noise in our measurements. In the other two methods, we combine either the measured surface gravity or the measured radius with theoretical WD mass-radius relations from the La Plata group. We find that both our mass distributions from surface gravity alone and radius alone agree well with that of \citet{Kepler_2019}, with the distributions from surface gravity being slightly narrower and from radius being slightly broader than the \citet{Kepler_2019} distribution. Based on these comparisons, we find that catalog users can use either WD masses measured from radii combined with theory or WD masses measured from surface gravities combined with theory in studies that do not require the measured WD mass to be independent of an assumed mass-radius relation. 

\indent In previous generations of SDSS, WDs were targeted serendipitously as quasar candidates, creating spatial and color selection biases not present in the SDSS-V sample of WDs, in which WDs were explicitly targeted. These differences in survey strategy translate into differences in the observed population of DA WDs, which we investigate in Sec. \ref{sec:SDSS_diffs}. We find different spatial and Gaia $G$-band magnitude distributions between SDSS-V and previous generations, as expected. There were also significant changes in hardware, with the survey moving away from traditional plug plates to a fiber positioning system, and changes in reduction pipelines, with new procedures for the treatment of skylines on the blue end of the spectrum, during the SDSS-V era. Such changes could cause the same WD observed in SDSS-V and in previous generations to have differences in measured spectroscopic parameters. On average the same WD observed in both in SDSS-V and previous SDSS generations has a measured apparent radial velocity that is $11.5$ km/s larger and a surface gravity that is $0.015$ dex smaller when using SDSS-V data compared to previous SDSS data. Differences of this size warrant a comprehensive investigation of measured physical parameters for WDs observed both in SDSS-V and previous generations, and this is left to future work.

\indent This work is primarily limited by spectrum SNR. We find that measured apparent radial velocities, surface gravities, and spectroscopic effective temperatures are unreliable for spectra with SNRs $<10$, and all spectroscopic parameters are best measured from spectra with SNRs of at least 20. Most SDSS WD spectra have SNRs $<10$, making it difficult to accurately measure the properties of these stars. Additionally, of all our measured parameters, our characterization of spectroscopic surface gravities has the most need for improvement. We find that the \citet{Tremblay_2019_wdtools} training data set for the random forest routine has a residual high $\log{g}$ bump and some clumpy structure in the $11,000$ - $13,000$ K temperature range, leading to similar structure in our own fits. We find that our mass distribution from surface gravities and theory is more sharply peaked than the \citet{Kepler_2019} distribution and our distribution from photometric radii and theory. This sharper peak may indicate that the random forest routine biases WD surface gravities towards the center of the WD mass distribution, where there is more training data available. Random forests are a relatively simple way to create a regression between the shape of the spectral lines and the labeled surface gravity and temperature of the star, so future work could improve the \texttt{wdtools} generative fitting pipeline as an alternative. The generative fitting pipeline uses a more sophisticated neural network approach but is currently limited by its continuum normalization routine. Another approach would be to test other machine learning methods to create a better regressions between line shapes and spectroscopic labels. In particular, XGBoost, a gradient boosted tree-based method, has become increasingly popular for these sorts of applications \citep{Rene_2023}. Additionally, improving the choice of training data set would increase the reliability of our surface gravity measurements in the high $\log{g}$ regime.

\indent Future work will focus on improving our catalog in order to apply these measurements to better our understanding of WD astrophysics and to search for new fundamental physics. SDSS-V is on-going and will continue to grow the number of DA WDs observed. The more spectra we can coadd for each WD, the higher the spectrum SNR and the better we can characterize the apparent radial velocity, surface gravity, effective temperature of each object. The catalog produced in this work has already been used to detect the temperature-dependence of the WD mass-radius relation \citep{Crumpler_2024}, and extensions of that analysis due to improvements in our catalog should be able to constrain theoretical models of the WD mass-radius relation as a function of temperature. Additionally, this measurement pipeline can be applied to the sub-exposures comprising SDSS spectra to search for apparent radial velocity variation and binary candidates (Adamane Pallathadka et al., in prep). As we continue to refine our understanding of DA WD astrophysics, this catalog will be able to be used by the theoretical physics community to place new constraints on dark matter models. One interesting avenue is to probe the dark matter field structure on small spatial scales using the ubiquity of DA WDs. In particular, if dark matter is an ultra-light boson with standard-model interactions, then the coherence length scale of the field is set by the mass of the dark matter particle and dark matter would produce variations in the WD mass-radius relation on this length scale. An analysis of spatial correlations in deviations from the WD mass-radius relation could unveil this characteristic length scale and detect the mass of ultra-light dark matter (Crumpler et al., in prep).

\section{Acknowledgements} \label{sec:acknow}

\indent N.R.C is supported by the National Science Foundation Graduate Research Fellowship Program under Grant No. DGE2139757. Any opinions, findings, and conclusions or recommendations expressed in this material are those of the author and do not necessarily reflect the views of the National Science Foundation. V.C. gratefully acknowledges a Peirce Fellowship from Harvard University. N.L.Z. acknowledges support by the JHU President’s Frontier Award and by the seed grant from the JHU Institute for Data Intensive Engineering and Science. S.A. was supported by the JHU Provost’s Undergraduate Research Award. C.B. acknowledges support from NSF grant AST-2307865.

\indent Funding for the Sloan Digital Sky Survey V has been provided by the Alfred P. Sloan Foundation, the Heising-Simons Foundation, the National Science Foundation, and the Participating Institutions. SDSS acknowledges support and resources from the Center for High-Performance Computing at the University of Utah. SDSS telescopes are located at Apache Point Observatory, funded by the Astrophysical Research Consortium and operated by New Mexico State University, and at Las Campanas Observatory, operated by the Carnegie Institution for Science. The SDSS web site is \url{www.sdss.org}.

\indent SDSS is managed by the Astrophysical Research Consortium for the Participating Institutions of the SDSS Collaboration, including the Carnegie Institution for Science, Chilean National Time Allocation Committee (CNTAC) ratified researchers, Caltech, the Gotham Participation Group, Harvard University, Heidelberg University, The Flatiron Institute, The Johns Hopkins University, L'Ecole polytechnique f\'{e}d\'{e}rale de Lausanne (EPFL), Leibniz-Institut f\"{u}r Astrophysik Potsdam (AIP), Max-Planck-Institut f\"{u}r Astronomie (MPIA Heidelberg), Max-Planck-Institut f\"{u}r Extraterrestrische Physik (MPE), Nanjing University, National Astronomical Observatories of China (NAOC), New Mexico State University, The Ohio State University, Pennsylvania State University, Smithsonian Astrophysical Observatory, Space Telescope Science Institute (STScI), the Stellar Astrophysics Participation Group, Universidad Nacional Aut\'{o}noma de M\'{e}xico, University of Arizona, University of Colorado Boulder, University of Illinois at Urbana-Champaign, University of Toronto, University of Utah, University of Virginia, Yale University, and Yunnan University.

\indent This work has made use of data from the European Space Agency (ESA) mission Gaia (\url{https://www. cosmos.esa.int/gaia}), processed by the Gaia Data Processing and Analysis Consortium (DPAC, \url{https://www. cosmos.esa.int/web/gaia/dpac/consortium}). Funding for the DPAC has been provided by national institutions, in particular the institutions participating in the Gaia Multilateral Agreement.

\textit{Software}: astropy \citep{Astropy_2013,Astropy_2018,Astropy_2022}

\appendix
\section{Appendix: SDSS-V Catalog Data Model} \label{sec:app1}

\startlongtable
\begin{deluxetable*}{lcl}
\tablecaption{SDSS-V Catalog Data Model\label{tab:sdssv}}
\tablehead{ \colhead{Column}  & \colhead{Units} & 
\colhead{Description}  } 
\renewcommand{\arraystretch}{0.9} 
\startdata
gaia\_dr3\_source\_id & - & Gaia DR3 source identifier \\
fieldid$^*$ & - & Field identifier \\
mjd$^*$ & - & Modified Julian date of observation \\
catalogid$^*$ & - & SDSS-V catalog identifier \\
snr$^*$ & - & Spectrum SNR \\
p\_da$^*$ & - & \texttt{SnowWhite} DA-type white dwarf probability \\
ra & deg & Right ascension \\
dec & deg & Declination \\
l & deg & Galactic longitude \\
b & deg & Galactic latitude \\
r\_med\_geo & pc & Median geometric distance from \citet{BailerJones_2021} \\
r\_lo\_geo & pc & 16th percentile of geometric distance from
\\&&\citet{BailerJones_2021} \\
r\_hi\_geo & pc & 84th percentile of geometric distance from
\\&&\citet{BailerJones_2021} \\
pmra & mas/yr & Proper motion in ra \\
pmra\_error & mas/yr & Error on proper motion in ra \\
pmdec & mas/yr & Proper motion in dec \\
pmdec\_error & mas/yr & Error on proper motion in dec \\
phot\_(g/bp/rp)\_mean\_flux & e-/s & Gaia $G$, $G_{\text{BP}}$, and $G_{\text{RP}}$-band mean fluxes \\
phot\_(g/bp/rp)\_mean\_flux\_error & e-/s & Error on $G$, $G_{\text{BP}}$, and $G_{\text{RP}}$-band mean fluxes  \\
phot\_(g/bp/rp)\_mean\_mag & mag & Gaia $G$, $G_{\text{BP}}$, and $G_{\text{RP}}$-band mean magnitudes on Vega scale\\
phot\_bp\_rp\_excess\_factor & - & Excess flux in Gaia $G_{\text{BP}}$/$G_{\text{RP}}$ photometry relative to $G$ band \\
no\_gaia\_phot & - & Flag indicating if WD lacks Gaia $G_{\text{BP}}$ or $G_{\text{RP}}$ mean fluxes \\
clean & - & SDSS clean photometry flag (1=clean, 0=unclean) \\
psf\_mag\_(u/g/r/i/z) & mag & SDSS PSF \textit{u}, \textit{g}, \textit{r}, \textit{i}, and \textit{z}-band magnitudes on the SDSS
\\&&scale \\
psf\_magerr\_(u/g/r/i/z) & mag & Error on SDSS PSF \textit{u}, \textit{g}, \textit{r}, \textit{i}, and \textit{z}-band magnitudes on
\\&&the SDSS scale \\
psf\_flux\_(u/g/r/i/z) & nanomaggies & SDSS PSF \textit{u}, \textit{g}, \textit{r}, \textit{i}, and \textit{z}-band fluxes  \\
psf\_fluxivar\_(u/g/r/i/z) & nanomaggies$^{-2}$ & Inverse variance of SDSS PSF \textit{u}, \textit{g}, \textit{r}, \textit{i}, and \textit{z}-band fluxes \\
mag\_ab\_(u/g/r/i/z) & mag & SDSS PSF \textit{u}, \textit{g}, \textit{r}, \textit{i}, and \textit{z}-band magnitudes on the AB
\\&&scale \\
magerr\_ab\_(u/g/r/i/z) & mag & Error on SDSS PSF \textit{u}, \textit{g}, \textit{r}, \textit{i}, and \textit{z}-band magnitudes on the
\\&&AB scale \\
no\_sdss\_phot & - & Flag indicating if WD lacks or has non-physical SDSS \textit{u}, \textit{r},
\\&&or \textit{z} magnitudes on the AB scale \\
rv\_(falcon/raddi/anguiano/ & km/s & WD apparent radial velocity in cross-matched published catalog \\
kepler)\\
e\_rv\_(falcon/raddi/anguiano/ & km/s & Error on published WD apparent radial velocity \\
kepler) \\
logg\_(raddi/anguiano/gentile/ & $\log$(cm/s$^2$) & WD surface gravity in cross-matched published catalog \\
koester/kepler) \\
e\_logg\_(raddi/anguiano/gentile/ & $\log$(cm/s$^2$) & Error on published WD surface gravity \\
koester/kepler)\\
teff\_(raddi/anguiano/gentile/ & K & WD effective temperature in cross-matched published catalog \\
/koester/kepler) \\
e\_reff\_(raddi/anguiano/gentile/ & K & Error on published WD effective temperature \\
koester/kepler) \\
radius\_(raddi/anguiano/ & $R_\odot$ & WD radius in cross-matched published catalog \\
gentile/kepler) \\
e\_radius\_(raddi/anguiano/ & $R_\odot$ & Error on published WD radius \\
gentile/kepler) \\
mass\_(raddi/anguiano/gentile/ & $M_\odot$ & WD mass in cross-matched published catalog \\
kepler) \\
e\_mass\_(raddi/anguiano/gentile/ & $M_\odot$ & Error on published WD mass \\
kepler) \\
(falcon/raddi/anguiano/gentile/ & - & Flag indicating whether WD is contained in cross-matched \\
koester/kepler)\_flag && published catalog\\
nspec\_coadd & - & Number of field-mjd-catalogid spectra corresponding to
\\&&unique Gaia DR3 source ID, all spectra are coadded to
\\&&create 1 spectrum per WD \\
snr\_coadd & - & Coadded spectrum SNR \\
rv\_corv\_(ind$^*$/coadd$^!$/mean) & km/s & Apparent radial velocity measured from each individual spectrum,
\\&& from each WD coadded spectrum, and from taking the weighted
\\&&mean of all high SNR individual spectrum radial velocities \\
e\_rv\_corv\_(ind$^*$/coadd/mean) & km/s & Measurement error on the apparent radial velocity\\
teff\_corv\_(ind$^*$/coadd) & K & \texttt{corv} effective temperature measured from each individual spectrum and
\\&& from each WD coadded spectrum, these measurements are not validated,
\\&& and the PRF parameters should be used instead\\
logg\_corv\_(ind$^*$/coadd) & $\log$(cm/s$^2$) & \texttt{corv} surface gravity measured from each individual spectrum and
\\&& from each WD coadded spectrum, these measurements are not validated,
\\&&and the PRF parameters should be used instead\\
e\_rv\_corv\_(ind$^*$/coadd$^!$)\_full & km/s & Full error (measured+systematic) on the apparent radial velocity \\
nspec\_mean\_rv\_corv & - & Number of high SNR spectra used to calculate weighted
\\&&mean apparent radial velocity \\
teff\_prf\_(ind$^*$/coadd/mean) & K & Effective temperature measured from each individual spectrum,
\\&& from each WD coadded spectrum, and from taking the weighted
\\&&mean of all high SNR individual spectrum temperatures \\
e\_teff\_prf\_(ind$^*$/coadd/mean) & K & Measurement error on the effective temperature\\
logg\_prf\_(ind$^*$/coadd$^!$/mean) & $\log$(cm/s$^2$) &  Surface gravity measured from each individual spectrum, from
\\&& each WD coadded spectrum, and from taking the weighted mean
\\&& of all high SNR individual spectrum surface gravities \\
e\_logg\_prf\_(ind$^*$/coadd/mean) & $\log$(cm/s$^2$) &  Measurement error on the surface gravity\\
e\_teff\_prf\_(ind$^*$/coadd)\_full & K & Full error (measured+systematic) on the effective temperature \\
e\_logg\_prf\_(ind$^*$/coadd$^!$)\_full & $\log$(cm/s$^2$) & Full error (measured+systematic) on the surface gravity \\
nspec\_mean\_teff & - & Number of high SNR spectra used to calculate weighted
\\&&mean effective temperature \\
nspec\_mean\_logg & - &  Number of high SNR spectra used to calculate weighted
\\&&mean surface gravity \\
av\_(lo/med/hi) & mag & Total extinction in Johnson–Cousins V-band at lo, med, or hi
\\&&geometric distance \\
(u/g/r/i/z)\_ext\_(lo/med/hi) & mag & Total extinction in SDSS \textit{u}, \textit{g}, \textit{r}, \textit{i}, or \textit{z}-bands at lo, med, or hi
\\&&geometric distance \\
gaia\_(g/bp/rp)\_ext\_(lo/med/hi) & mag & Total extinction in Gaia $G$, $G_{\text{BP}}$, or $G_{\text{RP}}$-bands at lo, med, or hi
\\&&geometric distance \\
(u/g/r/i/z)\_dered\_(lo/med/hi) & mag & Dereddened SDSS \textit{u}, \textit{g}, \textit{r}, \textit{i}, or \textit{z}-band magnitudes at lo, med, or hi
\\&&geometric distance \\
gaia\_(g/bp/rp)\_dered\_(lo/med/hi) & mag & Dereddened Gaia $G$, $G_{\text{BP}}$, or $G_{\text{RP}}$-band magnitudes at lo, med, or hi
\\&&geometric distance \\
phot\_used & - & Flag indicating whether SDSS or Gaia photometry was used to fit
\\&&WD parameters (1=SDSS, 2=Gaia) \\
phot\_radius\_(sdss/gaia)\_(lo/ & $R_\odot$ & Photometric radius measured at lo, med, or hi geometric distance\\
med/hi)& &with SDSS or Gaia photometry\\
e\_phot\_radius\_(sdss/gaia)\_(lo/ & $R_\odot$ & Measurement error on the photometric radius \\
med/hi)&&\\
phot\_teff\_(sdss/gaia)\_(lo/med/hi) & K & Photometric effective temperature measured at lo, med, or hi
\\&&geometric distance with SDSS or Gaia photometry\\
e\_phot\_teff\_(sdss/gaia)\_(lo/med/hi) & K & Measurement error on the photometric effective temperature \\
phot\_logg\_sdss\_(lo/med/hi) & $\log$(cm/s$^2$) & Photometric surface gravity measured at lo, med, or hi
\\&&geometric distance with SDSS photometry\\
e\_phot\_logg\_sdss\_(lo/med/hi) & $\log$(cm/s$^2$) & Measurement error on the photometric surface gravity \\
(sdss/gaia)\_logg\_flag\_(lo/med/hi) & - & Flag indicating whether the photometric fit 0=failed, 1=was
\\&&successful with variable $\log g$, 2=was successful with fixed $\log g=8$\\
phot\_redchi\_(sdss/gaia)\_(lo/med/hi) & - & Reduced $\chi^2$ on the photometric fit\\
radius\_phot\_(sdss/gaia) & $R_\odot$ & Final SDSS or Gaia photometric radius measurement, taken to be
\\&&the value at the median geometric distance \\
e\_radius\_phot\_(sdss/gaia) & $R_\odot$ & Error (measured+distance uncertainty) on the final SDSS or
\\&& Gaia photometric radius \\
teff\_phot\_(sdss/gaia) & K & Final SDSS or Gaia photometric effective temperature measurement,
\\&&taken to be the value at the median geometric distance \\
e\_teff\_phot\_(sdss/gaia) & K & Error (measured+distance uncertainty) on the final SDSS or
\\&& Gaia photometric effective temperature \\
phot\_err\_sdss & mag & Mean SDSS \textit{u}, \textit{r}, and \textit{z}-band magnitude error \\
phot\_err\_gaia & e-/s & Mean Gaia $G_{\text{BP}}$ and $G_{\text{RP}}$ flux error \\
radius\_phot$^!$ & $R_\odot$ & Final photometric radius measurement, prioritizing SDSS
\\&&photometry, taken to be the value at the median geometric distance \\
e\_radius\_phot & $R_\odot$ & Error (measured+distance uncertainty) on the final photometric
\\&&radius \\
teff\_phot$^!$ & K & Final photometric effective temperature measurement, prioritizing
\\&&SDSS photometry, taken to be the value at the median geometric
\\&&distance \\
e\_teff\_phot & K & Error (measured+distance uncertainty) on the final photometric
\\&&effective temperature \\
e\_radius\_phot\_full$^!$ & $R_\odot$ & Full error (measured+distance uncertainty+systematic) on the
\\&&final radius\\
e\_teff\_phot\_full$^!$ & K &  Full error (measured+distance uncertainty+systematic) on the
\\&&final effective temperature\\
mass\_rad\_logg & $M_\odot$ & Mass measured from coadded spectrum surface gravity and
\\&&final photometric radius \\
e\_mass\_rad\_logg & $M_\odot$ & Error on mass from surface gravity and radius \\
mass\_logg\_theory & $M_\odot$ & Mass measured from coadded spectrum surface gravity and
\\&& effective temperature, combined with La Plata models \\
e\_mass\_logg\_theory & $M_\odot$ & Error on mass measured from surface gravity and theory \\
mass\_rad\_theory$^!$ & $M_\odot$ & Mass measured from final photometric radius and 
\\&&effective temperature, combined with La Plata models\\
e\_mass\_rad\_theory$^!$ & $M_\odot$ & Error on mass measured from radius and theory \\
rv\_corv\_lsr & km/s & Measured coadded spectrum apparent radial velocity,
\\&&corrected to the LSR \\
rv\_corv\_asym\_corr & km/s & Measured coadded spectrum apparent radial velocity,
\\&&corrected to the LSR and for asymmetric drift,
\\&&only for WDs used in the \citet{Crumpler_2024} detection\\
asym\_corr & km/s & Applied asymmetric drift correction, only for WDs used in
\\&&the \citet{Crumpler_2024} detection \\
tempdep\_catalog\_flag & - & Flag indicating whether WD was used in the 
\\&&\citet{Crumpler_2024} detection\\
eta & - & Logarithm of the probability that the observed apparent radial
\\&&velocity variation is random noise, $\eta$ \\
ruwe & - & Gaia Renormalised Unit Weight Error (RUWE) \\
binary\_flag & - & Flag indicating whether the WD is a potential binary
\\&& 0= No evidence for binarity
\\&& 1= Evidence for binarity from apparent radial velocity variation
\\&& 2= Evidence for binarity from Gaia RUWE
\\&& 3= Evidence for binarity from both apparent radial velocity variation
\\&& and RUWE\\
speed\_lsr & km/s & WD total speed relative to the LSR \\
star\_pop\_flag & - &  Flag indicating whether WD likely belongs to the 
\\&& 0= Thick disk or uncertain designation
\\&& 1= Thin disk
\\&& 2= Halo
\enddata
\tablecomments{For each WD, there are sometimes multiple measurements of the same parameter with different methods. We flag each column containing the best-measurement for each unique WD with ``$^!$", and suggest that members of the community use these values. Columns with a ``$^*$" indicate that there is one unique value for each fieldID-mjd-catalogID spectrum. Otherwise, there is one value for each WD as designated by a unique Gaia DR3 source ID, and the value may occur several times in the table.}
\end{deluxetable*} 

\section{Appendix: Previous SDSS Catalog Data Model} \label{sec:app2}

\startlongtable
\begin{deluxetable*}{lcl}
\tablecaption{Previous SDSS Catalog Data Model\label{tab:esdss}}
\tablehead{ \colhead{Column}  & \colhead{Units} & \colhead{Description}  } 
\renewcommand{\arraystretch}{0.9} 
\startdata
gaia\_dr3\_source\_id & - & Gaia DR3 source identifier \\
plate$^*$ & - & SDSS plate identifier \\
mjd$^*$ & - & Modified Julian date of observation \\
fiber$^*$ & - & SDSS fiber identifier \\
spec\_avail  & - & Flag indicating where the spectrum can be accessed:
\\&&0= \url {https://dr18.sdss.org/sas/dr18/spectro/sdss/redux/26/spectra/lite/},
\\&&1= \url {https://dr18.sdss.org/sas/dr18/spectro/sdss/redux/103/spectra/lite/}, 
\\&&2= \url {https://dr18.sdss.org/sas/dr18/spectro/sdss/redux/104/spectra/lite/},
\\&&3= \url {https://dr18.sdss.org/sas/dr18/spectro/sdss/redux/v5_13_2/spectra/lite}/,
\\&&4= \url {https://dr18.sdss.org/sas/dr18/spectro/sdss/redux/v6_0_4/spectra/lite/},
\\&&5= \url {https://dr18.sdss.org/sas/dr18/spectro/sdss/redux/eFEDS/spectra/lite/},
\\&&6= \url {https://dr18.sdss.org/sas/dr18/spectro/boss/redux/eFEDS/spectra/lite/},
\\&&7= \url {https://dr18.sdss.org/sas/dr18/spectro/boss/redux/v6_0_4/spectra/lite/},
\\&&8= available through \texttt{SDSS.astroquery}\\
sdss\_dr  & - & Most recent published SDSS WD catalog containing this WD, all listed 
\\&&measurements are from this data release\\
rv\_sdss\_dr & km/s & WD apparent radial velocity in published SDSS catalog \\
e\_rv\_sdss\_dr & km/s & Error on published WD apparent radial velocity \\
teff\_(sdss\_dr/gentile) & K & WD effective temperature in published SDSS catalog or \citet{Gentile_2021} \\
e\_teff\_(sdss\_dr/gentile) & K & Error on published WD effective temperature \\
logg\_(sdss\_dr/gentile) & $\log$(cm/s$^2$) & WD surface gravity in published SDSS catalog or \citet{Gentile_2021} \\
e\_logg\_(sdss\_dr/gentile) & $\log$(cm/s$^2$) & Error on published WD surface gravity \\
teff\_1d\_sdss\_dr & K & WD effective temperature in published SDSS catalog, not corrected for 3D effects\\
e\_teff\_1d\_sdss\_dr & K & Error on published WD effective temperature, not corrected for 3D effects\\
logg\_1d\_sdss\_dr & $\log$(cm/s$^2$) & WD surface gravity in published SDSS catalog, not corrected for 3D effects\\
e\_logg\_1d\_sdss\_dr & $\log$(cm/s$^2$) & Error on published WD surface gravity, not corrected for 3D effects \\
radius\_(sdss\_dr/gentile) & $R_\odot$ & WD radius in published SDSS catalog or or \citet{Gentile_2021}\\
e\_radius\_(sdss\_dr/gentile) & $R_\odot$ & Error on published WD radius \\
radius\_1d\_sdss\_dr & $R_\odot$ & WD radius in published SDSS catalog, not corrected for 3D effects \\
e\_radius\_1d\_sdss\_dr & $R_\odot$ & Error on published WD radius, not corrected for 3D effects \\
mass\_(sdss\_dr/gentile) & $M_\odot$ & WD mass in published SDSS catalog or \citet{Gentile_2021}\\
e\_mass\_(sdss\_dr/gentile) & $M_\odot$ & Error on published WD mass \\
mass\_1d\_sdss\_dr & $M_\odot$ & WD mass in published SDSS catalog, not corrected for 3D effects \\
e\_mass\_1d\_sdss\_dr & $M_\odot$ & Error on published WD mass, not corrected for 3D effects
\enddata
\tablecomments{The following columns are included in the previous SDSS catalog, but were not redescribed as they have already been discussed in Appendix \ref{sec:app1}:\newline 
snr$^*$, ra, dec, l, b, r\_med\_geo, r\_lo\_geo, r\_hi\_geo, pmra, pmra\_error, pmdec, pmdec\_error, 
phot\_(g/bp/rp)\_mean\_flux, phot\_(g/bp/rp)\_mean\_flux\_error, phot\_(g/bp/rp)\_mean\_mag,
phot\_bp\_rp\_excess\_factor,no\_gaia\_phot, clean, psf\_mag\_(u/g/r/i/z), psf\_magerr\_(u/g/r/i/z),
psf\_flux\_(u/g/r/i/z), psf\_fluxivar\_(u/g/r/i/z), mag\_ab\_(u/g/r/i/z), magerr\_ab\_(u/g/r/i/z), no\_sdss\_phot, 
snr\_coadd, nspec\_coadd, rv\_corv\_(ind$^*$/coadd$^!$/mean), e\_rv\_corv\_(ind$^*$/coadd/mean), 
teff\_corv\_(ind$^*$/coadd/mean), 
\newline logg\_corv\_(ind$^*$/coadd/mean), e\_rv\_corv\_(ind$^*$/coadd$^!$)\_full, 
nspec\_mean\_rv\_corv, teff\_prf\_(ind$^*$/coadd/mean),
\newline e\_teff\_prf\_(ind$^*$/coadd/mean), logg\_prf\_(ind$^*$/coadd$^!$/mean), e\_logg\_prf\_(ind$^*$/coadd/mean), e\_teff\_prf\_(ind$^*$/coadd)\_full,
\newline e\_logg\_prf\_(ind$^*$/coadd$^!$)\_full, nspec\_mean\_teff, nspec\_mean\_logg, av\_(lo/med/hi), 
(u/g/r/i/z)\_ext\_(lo/med/hi), 
\newline gaia\_(g/bp/rp)\_ext\_(lo/med/hi), (u/g/r/i/z)\_dered\_(lo/med/hi), 
gaia\_(g/bp/rp)\_dered\_(lo/med/hi), phot\_used, \newline phot\_radius\_(sdss/gaia)\_(lo/med/hi), 
e\_phot\_radius\_(sdss/gaia)\_(lo/med/hi), phot\_teff\_(sdss/gaia)\_(lo/med/hi), 
\newline e\_phot\_teff\_(sdss/gaia)\_(lo/med/hi), phot\_redchi\_(sdss/gaia)\_(lo/med/hi), 
phot\_logg\_sdss\_(lo/med/hi),
\newline e\_phot\_logg\_sdss\_(lo/med/hi), (sdss/gaia)\_logg\_flag\_(lo/med/hi), 
radius\_phot\_(sdss/gaia), e\_radius\_phot\_(sdss/gaia), 
\newline teff\_phot\_(sdss/gaia), e\_teff\_phot\_(sdss/gaia), 
phot\_err\_(sdss/gaia), radius\_phot$^!$, e\_radius\_phot, teff\_phot$^!$, e\_teff\_phot,
\newline e\_radius\_phot\_full$^!$, 
e\_teff\_phot\_full$^!$, mass\_rad\_logg, e\_mass\_rad\_logg, mass\_logg\_theory, e\_mass\_logg\_theory, 
mass\_rad\_theory$^!$, e\_mass\_rad\_theory$^!$, rv\_corv\_lsr, rv\_corv\_asym\_corr, asym\_corr, tempdep\_catalog\_flag, eta, ruwe, binary\_flag, speed\_lsr, star\_pop\_flag
}
\tablecomments{For each WD, there are sometimes multiple measurements of the same parameter with different methods. We flag each column containing the best-measurement for each unique WD with ``$^!$", and suggest that members of the community use these values. Columns with a ``$^*$" indicate that there is one unique value for each plate-mjd-fiberID spectrum. Otherwise, there is one value for each WD as designated by a unique Gaia DR3 source ID, and the value may occur several times in the table.}
\end{deluxetable*} 

\clearpage

\bibliographystyle{aasjournal}
\bibliography{bibtex}{}

\end{document}